\documentclass[fleqn,usenatbib]{mnras}

\usepackage{newtxtext,newtxmath}

\usepackage[T1]{fontenc}
\usepackage{ae,aecompl}

\usepackage{graphicx}	
\usepackage{amsmath}	
\usepackage{amssymb}	
\usepackage{xspace}

\def\mpcoh{\,h^{-1}{\rm Mpc}}
\def\kpcoh{\,h^{-1}{\rm kpc}}
\def\msolaroh{\,h^{-1}M_\odot}
\DeclareMathOperator\erf{erf}
\DeclareMathOperator\erfc{erfc}
\newcommand\mthod{\texttt{MTHOD}}
\newcommand{\mufasa}{\mbox{{\sc \small Mufasa}}\xspace}
\newcommand{\illustris}{\mbox{{\sc \small Illustris}}\xspace}

\title[Multi-Tracer HOD]{Multi-tracer extension of the halo model: probing quenching and conformity in eBOSS}

\author[Alam et al.]{%
Shadab Alam,$^{1}$\thanks{E-mail: salam@roe.ac.uk}
John A. Peacock,$^{1}$
Katarina Kraljic,$^{1}$
Ashley J. Ross,$^{2}$ 
and \and Johan Comparat$^{3}$
\\
$^{1}$ Institute for Astronomy, University of Edinburgh, Royal Observatory, Blackford Hill, Edinburgh, EH9 3HJ , UK \\
$^{2}$ Center for Cosmology and AstroParticle Physics, The Ohio State University, 191 West Woodruff Avenue, Columbus, Ohio 43210, USA\\
$^{3}$ Max-Planck-Institut f\"{u}r extraterrestrische Physik (MPE), 
Giessenbachstrasse 1, D-85748 Garching bei M\"unchen, Germany\vspace{-6pt}\\
}

\pubyear{2020}

\begin{document}
\label{firstpage}
\pagerange{\pageref{firstpage}--\pageref{lastpage}}
\maketitle

\begin{abstract}
We develop a new Multi-Tracer Halo Occupation Distribution (\mthod)
framework for the galaxy distribution and apply it to the extended
Baryon Oscillation Spectroscopic Survey (eBOSS) final data between
$z=0.7-1.1$. We obtain a best fit \mthod\, for each tracer and describe
the host halo properties of these galaxies. The mean halo masses for LRGs,
ELGs and QSOs are found to be $1.9 \times 10^{13} \msolaroh$, $1.1
\times 10^{12} \msolaroh$ and $5 \times 10^{12} \msolaroh$ respectively
in the eBOSS data. We use the \mthod\, framework to create mock galaxy
catalogues and predict auto- and cross-correlation functions for all
the tracers. Comparing these results with data, we investigate
galactic conformity, the phenomenon whereby the properties of neighbouring galaxies are mutually correlated  in a manner that is not captured by the basic halo model. We detect \textsl{1-halo} conformity at
more than 3$\sigma$ statistical significance, while obtaining upper
limits on \textsl{2-halo} conformity. We also look at the environmental
dependence of the galaxy quenching efficiency and find that halo mass
driven quenching successfully explains the behaviour in high density
regions, but it fails to describe the quenching efficiency in low density regions. In particular, we show that the quenching efficiency in low density filaments is higher in the observed data, as compared to the prediction of the \mthod\ with halo mass driven quenching. The mock galaxy catalogue constructed in this paper is publicly available on this website\footnotemark.
\end{abstract}
\footnotetext{\url{https://www.roe.ac.uk/~salam/MTHOD/}}



\section{Introduction}
Galaxies are fundamental building blocks of the Universe, and
their large-scale spatial distribution is an important source
of cosmological information. However, a full understanding of the
formation of these objects within the cosmic web of
large-scale structure is challenging, owing to the number of astrophysical
processes that must be modelled and the wide dynamic range of scales on
which these operate.
In principle the dark matter distribution should give a direct and precise answer
to where galaxies should form, given that the dominant physical process
is gravitational collapse of matter. Therefore, one can attempt to predict the
locations of galaxies in the Universe purely by studying dark matter
dynamics, which is a much simpler task in the $\Lambda$CDM model than performing calculations
that include full baryonic physics. 

This approach has led to the widely-used halo model
\citep[HOD:][]{Seljak2000,Peacock2000,Benson2000,White2001,Berlind2002,Cooray2002},
which establishes the connection between dark matter haloes and
galaxies. It has been highly successful for magnitude limited sample of
bright galaxies in past galaxy redshift surveys
\citep[e.g.][]{Eisenstein2011,gama2015}. The current generation of surveys
\citep[e.g. eBOSS:][]{2016AJ....151...44D} and forthcoming surveys
\citep[e.g. DESI:][]{2016arXiv161100036D} are in contrast
targeting deeper samples and in particular selecting galaxies with special
photometric properties. This brings a new challenge in terms of understanding
the special locations that these galaxies might inhabit within the dark matter distribution,
so that we can use them as tracers in order to obtain
unbiased cosmological constraints. But this also offers a great
opportunity in using these galaxies to understand certain aspects of
galaxy formation physics, and to enhance cosmological constraints by
exploiting information contained in the cross-correlations between
the different populations. This goal requires the development of methods that not only predict the locations of galaxies within the population of dark matter haloes, but which can also map their photometric
properties. There have been several attempts in the past to extend the
halo model in order to predict the photometric properties of galaxies \citep[e.g.][]{2002MNRAS.332..697S,2006MNRAS.365..842C,2009MNRAS.392.1080S,2018MNRAS.481.5470X}.

Galaxies exhibit a wide range of properties with rather diverse behaviours. The colour distribution of galaxies shows bi-modal behaviour representing largely red and blue population \citep[e.g.][]{Strateva2001,Baldry2004,Balogh2004}.
More massive galaxies mostly tend to be red spheroids, while less massive galaxies tend to be star-forming discs \citep[e.g.][]{kauffmann2003b}. For the purpose of cosmological surveys, one can conveniently think about three kinds of galaxies as follows:
\begin{itemize}
    \item Quenched galaxies: The sub-population of galaxies with
      effectively no active star formation. Such galaxies appear red
      in colour and are generally massive and old. They generally 
      live at the centres of massive haloes or in high density
      regions. There are several processes thought to be responsible
      for `quenching', i.e.
      the cessation of star formation (e.g. cold gas stripping, harassment, strangulation or starvation). Galaxy redshift surveys such as eBOSS  and DESI typically select
      a specific sub-sample of quenched
      galaxies, known as Luminous Red Galaxies (LRG), based on
      photometric observations (see Section~\ref{sec:data-LRG} for the details of eBOSS LRG selection).
    \item Star-forming galaxies: The sub-population of galaxies with
      active star formation. Such galaxies typically appear blue and
      generally avoid high densities or the centres of the
      most massive haloes. They are also expected to be
      predominantly low mass galaxies. The complex interplay
      between density and tidal environment makes it more difficult to
      predict the locations of such galaxies. eBOSS and DESI target
      a very specific sub-sample of star forming galaxies with
      high emission in OII 3727\AA\ line flux -- Emission Line galaxies (ELG; see Section~\ref{sec:data-ELG} for the details of eBOSS ELG selection).
    \item AGN: Galaxies with strong emission from their central black
      hole region known as Active Galactic Nuclei (AGN). The importance
      of environment and baryonic processes during their formation
      makes the ability to predict their location within dark matter
      distribution challenging. eBOSS and DESI target a
      specific subset of such AGN that are especially bright in the optical bands,
      also known as Quasi Stellar Objects (QSOs; see Section~\ref{sec:data-QSO} for the details of eBOSS QSO selection).
\end{itemize}

The current and future surveys of these three types of galaxies have
great potential to measure whether these different kinds
of galaxies have any special clustering properties, and thus to give insights into the
physical processes involved in their formation. At the same time, more
complicated formation mechanisms introduce astrophysical
systematics in the clustering properties of these galaxies, which need
to be accounted for in order to infer unbiased cosmological
parameters. The halo model assumes that halo mass can predict the location
of different kinds of galaxies within the dark matter field. One can argue
that non-linear evolution of both dark matter and baryons can add
additional features in the relation between dark matter and galaxies. For
example large-scale tidal fields can affect the accretion rate of haloes
\citep{Musso2018MNRAS.tmp..187M}, with smaller haloes around filaments possibly growing more slowly
\citep{Hahn2009MNRAS.398.1742H,Castorina2016arXiv161103619C,Borzyszkowski2017MNRAS.469..594B}. 
In theoretical models, galaxy
quenching is assumed to be driven mainly by halo mass
\citep[e.g.][]{dekel2006,cattaneo2006, correa2018}; this concept is
successful in describing galaxy clustering and weak
gravitational lensing \citep{zu2016}. In principle, for galaxy
evolution, the cosmic web can also play a significant role by directing cold
gas flows along filaments \citep{keres2005}, thus providing material for
star formation to galaxies along filaments leading to suppression of
quenching \citep{kleiner2017}. It is also important to note that the central black holes in galaxies grow with galaxies. Therefore, a massive black hole can initiate feedback which can regulate the gas causing quenching. Such process can have impact on galaxy properties beyond halo mass.

In this paper we develop a halo model called Multi-Tracer Halo
Occupation Distribution (\mthod\, hereafter), which extends the standard
HOD to include the simultaneous treatment of multiple kinds of galaxies. The general statistical framework we present is extremely useful in practical modelling of current and future
surveys. This will allow us to interpret cross-correlation measurements and test the halo model in ways not possible with a single-tracer model. We apply the \mthod\, framework to eBOSS data and show its potential. 
In principle the vanilla version of the model contains minimal
astrophysical information on galaxy clustering. Our approach is always to
work with the simplest ansatz unless it fails through the data showing a
signature of additional physics. We test this model by making detailed
comparison with data and show that the fiducial model indeed fails to
reproduce observations exactly. We then develop a possible extension of the fiducial
model and highlight the insight this gives us regarding galaxy
formation physics. We do not perform cosmological parameter analysis here; this
will be addressed in future studies.

In particular, we focus on galactic conformity, which is a phenomenon whereby
the properties of neighbouring galaxies are mutually
correlated in a manner that is not captured by the basic halo model. 
Depending on the spatial scale considered, two regimes can
be distinguished: the so-called \textsl{1-halo} conformity, when the
neighbouring galaxies belong to the same halo, and \textsl{2-halo}
conformity, extending to scales beyond the virial radius of
haloes. Galactic conformity was first identified by
\cite{Weinmann2006} as an intriguing correlation between the
star-formation properties of central galaxies and those of their
satellites, with the fraction of red, passive satellites around a red,
passive central being significantly higher than around a blue, active
central. Since then, this \textsl{1-halo} conformity has been
confirmed by several authors both in the local Universe
\cite[e.g.][]{Kauffmann2010,WW2012,Kauffmann2013,Robotham2013,Phillips2014,Knobel2015,Treyer2018}
and at higher redshifts \cite[][]{Hartley2015,Kawin2016,Berti2017}. A galactic conformity type signal was also reported in \cite{2009MNRAS.399..878R} which used halo model and required early and late type galaxies to be in different halos.
Beyond this, a detection of \textsl{2-halo} conformity was reported 
by \cite{Kauffmann2013}; however, this result is still debated as
possibly driven by selection bias \citep{Sin2017,Lacerna2018,Tinker2018}. Interestingly \cite{calderon2017} reported no signature of \textsl{1-halo} but a robust detection of \textsl{2-halo} galactic conformity at low redshift in SDSS DR7 data. 
Galactic conformity is usually identified in observations
by first identifying central and satellite galaxies that are prone to
systematic errors. Our method avoids the need for such identification and
works at the population level to constrain this effect.

The paper is organised as follows. First, we introduce the details of
the \mthod\, model in Section 2, then we describe the eBOSS data used in
Section 3, followed by the description of the measurement and systematics
in Section 4. We describe the details of the analysis in Section 5 and present the results
of applying \mthod\, in Section 6. We finish by summarising our results
with discussion of their implication in Section 7.

\section{Data}
\label{sec:data}

\begin{figure}
    \centering
    \includegraphics[width=0.48\textwidth]{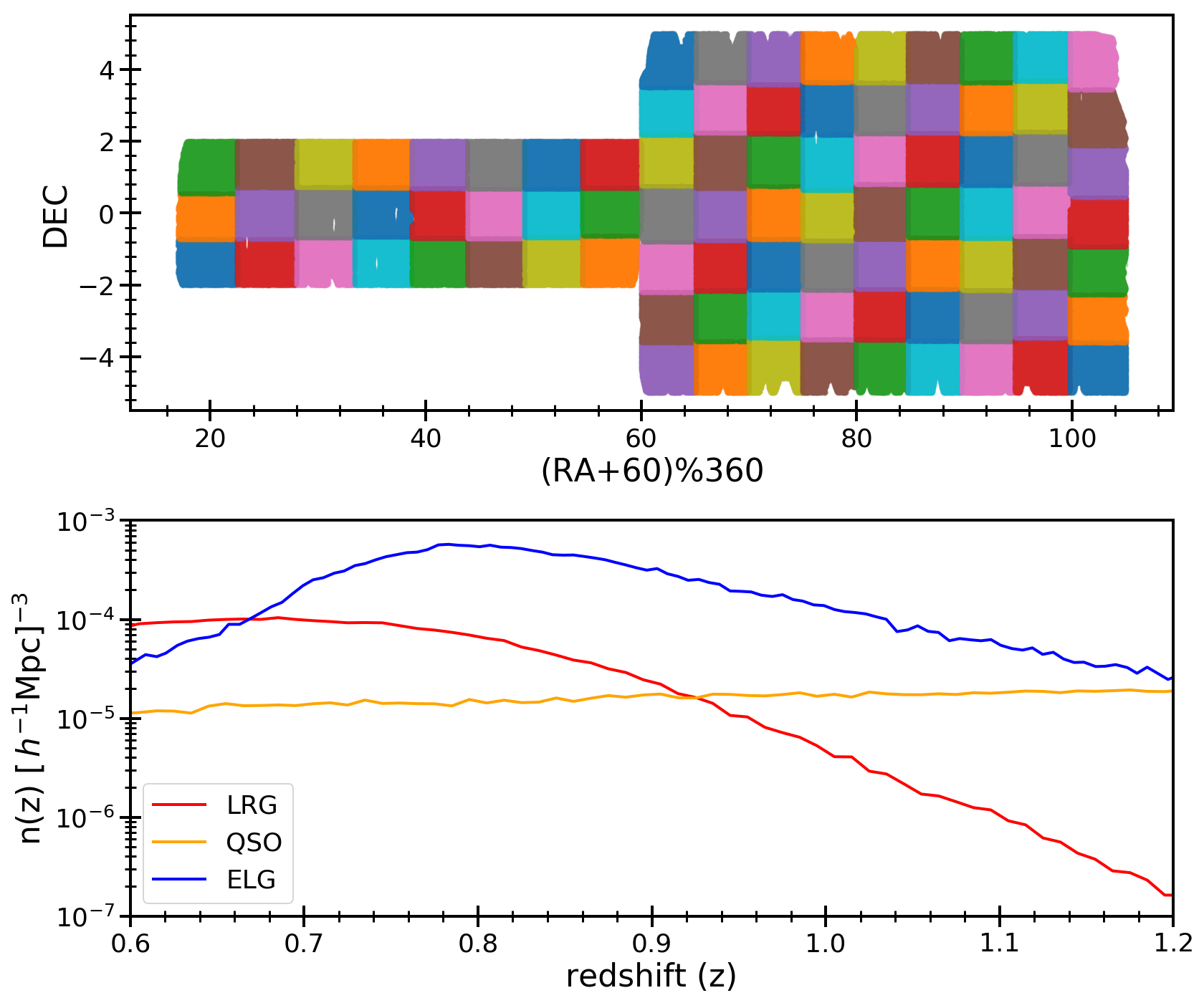}
    \caption{The sky coverage and number density distribution of our sample used in this paper. The top panel shows the sky coverage of the sub-sample of eBOSS data with different colours representing the jackknife realisations. The bottom panel shows the number density distribution of each tracer in the sample.}
    \label{fig:coverage}
\end{figure}

We are using data obtained by the extended Baryon Oscillation
Spectroscopic (eBOSS) survey \citep{Dawson2013}. The eBOSS project is one of
the programmes within the wider 5-year Sloan Digital Sky Survey-IV
\citep[SDSS-IV][]{2017AJ....154...28B} using the 2.5m Sloan Telescope
\citep{Gunn2006} and BOSS spectrograph \citep{Smee2013}. The eBOSS sample
consists of four different types of tracers, namely Luminous Red
Galaxies (LRG); Emission Line Galaxies (ELG); Quasi-Stellar Objects
(QSO); and Lyman Alpha Forest. We are using a subset of the eBOSS samples
that covers the redshift range from 0.7 to 1.1, where all of the three tracers of
interest, namely LRG, ELG and QSOs, overlap. The overlap region can be
used to study these tracers with cross-correlations and this results in
dense enough galaxy samples to probe the underlying dark matter
distribution through the combined samples. We use intermediate versions of the data release 16 (DR16) catalogues produced by the eBOSS collaboration (eBOSS collaboration in prep.; Raichoor et al. in prep). Any changes between the version we have used and the final versions are expected to be minor and mainly affect the results at large scales. We now briefly describe the relevant aspects of eBOSS sample selection.

\subsection{LRG selection}
\label{sec:data-LRG}
The LRGs are expected to be the brightest and reddest galaxies living in
massive dark matter haloes, and hence they have a high bias parameter. The eBOSS LRGs are
selected from SDSS imaging data \citep{Gunn2006, 2009ApJS..182..543A}
in combination with infrared photometry from WISE
\citep{2010AJ....140.1868W, 2014arXiv1410.7397L} using the following
target selection rules:
\begin{align}
    r-i & >0.98, \label{LRG1} \\
    r-W1 & > 2(r-i), \label{LRG2} \\
    i-z & >0.625, \label{LRG3}
\end{align}
where $r,i$ and $z$ are the model magnitudes of SDSS photometric bands
and $W1$ refers to the WISE photometry in the 3.4 micron channel. The
selections in equations \ref{LRG1}, \ref{LRG2} and \ref{LRG3} allow us
to pick out the desired redshift range, to reduce stellar contamination and to reduce
interlopers from below redshift 0.6 respectively. The details of how
these rules were derived and additional considerations are discussed
in \citet{2015ApJ...803..105P,2016ApJS..224...34P}.

\subsection{ELG selection}
\label{sec:data-ELG}
The ELGs are selected based on high OII flux and are expected to be star
forming galaxies that are typical of the population at high redshift. An earlier study about ELG selection with SDSS infrastructure was performed by \cite{2013MNRAS.428.1498C,2013MNRAS.433.1146C} and a pilot survey of ELGs testing different target selection algorithms is reported in \cite{2016A&A...592A.121C}. The ELG sample in eBOSS is selected
from the DECAM Legacy survey \cite[DECaLS:][]{2019AJ....157..168D}. The
target selection rules for ELGs in the North Galactic Cap (NGC) and South Galactic Cap (SGC) are slightly
different due to the availability of deeper data in the SGC. We use only the SGC region of the ELG sample for cross-correlations studies due to the
overlap with other tracers, hence we summarise only the SGC selection
here.  The ELG selection has two parts, the first of which is to select
star forming galaxies corresponding to the OII emission using the following
criterion:
\begin{equation}
        21.825  < g < 22.825  .
\end{equation}
The second rule for ELG selection is given by the following criteria, which preferentially select galaxies around redshift 1.0:
\begin{align}
    -0.068 (r-z) + 0.457 & < g-r < 0.112(r-z)+0.773,\\
    0.218 (g-r) + 0.571 & < r-z < -0.555(g-r) + 1.901,
\end{align}
where $g,r,z$ are the observed magnitudes in the DECaLS $g,r$ and $z$
photo-metric bands. More details of how these rules were derived and
additional considerations are discussed in \cite{2017MNRAS.471.3955R}.

\subsection{QSO selection}
\label{sec:data-QSO}
The density requirements of the other targets in the eBOSS sample restrict the
density of the eBOSS QSO sample. \cite{2015ApJS..221...27M} describe in
detail all the requirements and how the QSO sample is selected. First
a super-sample of QSOs is selected from the SDSS imaging with either
$g<22$ or $r<22$ and $i_f>17$, where $g$ and $r$ are the psf
magnitudes of the SDSS photometric bands and $i_f$ is the FIBER2MAG. This
super-sample is passed through the XDQSOz algorithm
\citep{2012ApJ...749...41B}, which assigns a probability for each object
being a QSO in a given redshift range using the photometric flux in
$ugriz$. The eBOSS sample uses a probabilistic cut of $P_{\rm  QSO}(z>0.9)>0.2$. 
There is also an infrared cut used to remove
stellar contamination with $m_{\rm opt}- m_{\rm WISE} \geq (g-i) +3$,
where $m_{\rm opt}$ and $m_{\rm WISE}$ are the optical and WISE
magnitudes respectively given by equations 1 and 2 in
\cite{2015ApJS..221...27M}.

We show the sky coverage of the eBOSS sub-sample, where all three tracers
are observed, in the top panel of Figure \ref{fig:coverage}. The
number density distribution of the three tracers used in this paper
are shown in the bottom panel of Figure \ref{fig:coverage}. We note
that around redshift 0.86 the number density of LRG, ELG and QSO
are $10^{-4}$, $4 \times 10^{-4}$ and $2 \times 10^{-5}
\smash{\left[\mpcoh\right]^{-3}}$, respectively. The redshift cut between 0.7 and 1.1 is chosen to have ELG number density above $10^{-4} \smash{\left[\mpcoh\right]^{-3}}$. The LRG number density drops sharply above redshift 0.9, this means that the cross-correlation signal of LRG is dominated by galaxies at lower redshift. We have looked at results by applying redshift cut between 0.7 and 0.9 and found the results are consistent with our fiducial redshift cut.

\subsection{Random catalogues and systematic weights}
Each of the large scale structure catalogues requires a corresponding random catalogue representing the sampling of the volume by the tracer. This is important to account for effects of survey geometry, survey mask, observational conditions, instrument efficiency etc. over the period of the full survey. The details of how these random catalogues are generated is described in \cite{ross20} for LRGs \& QSOs and \cite{anand20} for the ELG sample. The photometric galaxy sample on which target selection is applied for tracers shows correlation with various galactic maps such as stellar density, extinction coefficient, airmass, seeing and depth in photometric observations. The correlation of target density with such systematic maps is derived and corrected using weights described in \cite{ross20} and \cite{anand20}.

\section{Multi-Tracer HOD (\mthod)}
\label{sec:model}

We develop a generalised Halo Occupation Distribution model describing populations of multiple tracers.
The model assumes that all galaxies form and reside within dark matter
haloes, with properties dictated primarily by the mass of the host halo.
We further make the common assumption that any dark matter halo can have two types
of galaxies -- central and satellite. We describe the detailed model formulation for central
and satellite galaxies below.

\subsection{Central galaxies}
\label{sec:cengal}
The fiducial model assumes that the probability of having a central
galaxy of a given type is a function of halo mass only. The total probability for a halo to host a central galaxy is given by the summed probabilities of a central galaxy over all the tracers and is written as follows:
\begin{equation}
    p_{\rm cen}^{\rm tot}(M_{\rm h})=\sum_{tr \epsilon TR} p_{\rm cen}^{\rm tr}(M_{\rm h}),
\end{equation}
where the sum goes over all tracers in the list $TR=\left\{ \rm
LRG,QSO,ELG \right\}$. This equation requires a constraint of $p_{\rm
  cen}^{\rm tot}\leq1$ for any halo mass. This also assumes that the central
galaxy probabilities of different types of tracers are independent of
each other. This is a major assumption of the fiducial model and will
be discussed further in the later sections.

The LRG sample is assumed to inhabit largely massive haloes and
effectively $p_{\rm cen}^{\rm LRG}$ has been modelled successfully with a error function
given in equation \ref{eq:Nerrf} which is essentially saying that the
LRG sample central galaxy has a cut-off halo mass ($M_{\rm c}$) and a
dispersion ($\sigma_M$) in mass-to-light ratio:
\begin{equation}
    p_{\rm cen}^{\rm LRG}=\left< N_{\rm erf}(M_{\rm h}) \right> = \frac{1}{2} p_{\rm max} \erfc\left(\frac{\log_{10}{ M_{\rm c}}-\log_{10}{M_{\rm h}}}{\sqrt{2}  \log_{10}(e) \sigma_M}\right) .
    \label{eq:Nerrf}
\end{equation}
Where the parameter $p_{\rm max}$ controls the saturation level of the occupation probability and has been set to unity in the past, implying that the most massive haloes are guaranteed to possess a central galaxy.
Alternatively, Halo Abundance Matching (HAM) has also been used to model clustering of LRGs \citep{2016MNRAS.461.3421F}.
The statistics of central QSOs are harder to parameterize due to the
intrinsic complications of AGN physics and also practical issues of the optical selection
function. In the past, QSO central galaxies have been modelled using an
error function model \citep[e.g.][]{2012ApJ...755...30R}, but this is one of the major uncertainties regarding the halo
mass distribution of optically selected QSOs. For example, one can argue that LRGs have massive black holes and hence could host all QSOs with some duty cycle for black hole activity. Because the number of QSOs
is very small compared to other classes of galaxies and for the lack of any clearly preferred
alternative, we adopt the error function form given in equation
\ref{eq:Nerrf} to model the central galaxy probability for QSOs. We
will later discuss the implications of this assumption. How one can
learn about AGN physics by comparing different choices concerning central
galaxy occupation for QSOs will be studied in a follow-up paper (Alam et al. in prep.).
For a semi-analytic approach to model AGN see \cite{2019MNRAS.487..275G}.

\begin{figure}
    \centering
    \includegraphics[width=0.48\textwidth]{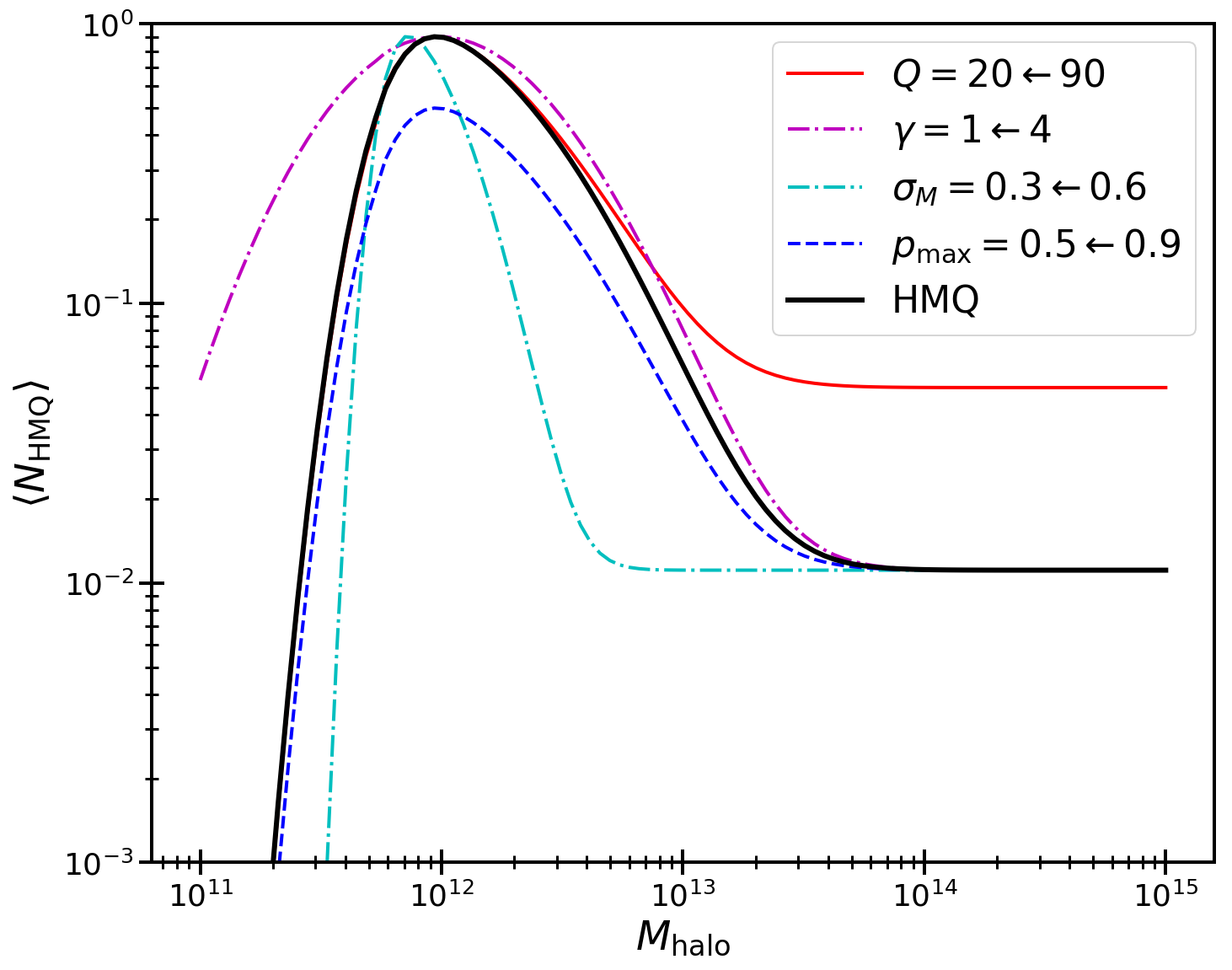}
    \caption{ELG occupation probability for  the Halo mass quenching model with an illustration of the effect of individual parameters. The black solid line shows fiducial model. The red solid, magenta dashed-dotted, cyan dashed-dotted and blue dashed lines show the impact of the parameters $Q$, $\gamma$, $\sigma_M$ and $p_{\rm max}$ respectively, when changed from the fiducial values. The legend also displays the values of parameters used in this illustration. }
    \label{fig:HMQ-illustration}
\end{figure}

Emission line galaxies (ELGs) emit strongly in OII 3737\AA, caused by active
star formation. Such star formation processes are inefficient at the
centres of massive haloes, reflecting the absence of cold gas. This
suggests that an error function model for ELG central galaxies may not be
realistic. \cite{2015A&A...575A..40C} suggested a semi-analytic model of galaxy formation based on the study of luminosity function of OII bright galaxies. \cite{2018MNRAS.474.4024G} have used this model to study ELGs and suggested that the central probability of ELGs should be reduced for high mass haloes as
shown in their Figure 9. Therefore, we propose the HMQ (High Mass Quenched) model for the central probability of ELGs with a skewed Gaussian function, as follows:
\begin{align}
    \left< N_{\rm HMQ}(M_{\rm h}) \right> &=  2 A \phi(M_{\rm h}) \Phi(\gamma M_{\rm h})  + & \nonumber \\  
    \frac{1}{2Q} & \left[1+\erf\left(\frac{\log_{10}{M_h}-\log_{10}{M_c}}{0.01}\right) \right],  \label{eq:NHMQ}\\
\phi(x) &=\mathcal{N}(\log_{10}{ M_c},\sigma_M), \label{eq:NHMQ-phi}\\
\Phi(x) &= \int_{-\infty}^x \phi(t) \, dt = \frac{1}{2} \left[ 1+\erf \left(\frac{x}{\sqrt{2}} \right) \right], \label{eq:NHMQ-Phi}\\
A &=\frac{p_{\rm max}  -1/Q }{\max(2\phi(x)\Phi(\gamma x))}.
\label{eq:NHMQ-A}
\end{align}
The effect of various parameters on the HMQ occupation function is illustrated in Figure~\ref{fig:HMQ-illustration}. The black line shows the fiducial model and each coloured line illustrates the impact of one parameter with details in the legend. The parameter $M_c$ is the cut-off mass of ELG centrals impacting the location of the peak in occupation probability and not shown in Figure~\ref{fig:HMQ-illustration} to avoid clutter. $Q$ sets the quenching efficiency for high mass haloes; a larger value
of $Q$ implies more efficient quenching as shown with a red solid line. The function $\phi(M_h)$ is the normal distribution given in equation~\ref{eq:NHMQ-phi} and $\Phi(M_h)$ is the cumulative density function of $\phi(M_h)$ given in equation~\ref{eq:NHMQ-Phi}. These two functions depend on the parameters $\gamma$ controlling the skewness as shown by magenta dashed-dotted line and $\sigma_M$ controlling the width illustrated with cyan dashed-dotted line. The parameter $A$ sets the overall formation efficiency of ELGs given in equation~\ref{eq:NHMQ-A} and depends on $p_{\rm max}$, which is illustrated with blue dashed line. 
We will also consider a more standard error function model
for ELG central galaxy probabilities in order to compare the
differences between the two cases. Note that the error function model for ELG central is non-physical and included only for the purpose of comparison.

Once we define the occupation recipe for central galaxies, we need to
assign their positions and velocities so that the redshift-space
clustering can be compared with data. The central galaxies in the
fiducial model are placed at the centre of the dark matter haloes and
the velocities given to them are the halo core velocities which are
computed using the particles within $10\%$ of the virial
radius of the centre of the halo. In principle, one could imagine that
the central galaxy could either deviate from the centre of the halo or
have velocity bias. 
Such variations will be considered in the future (Alam et al. in prep.) as the current data do not allow such an
extra freedom to be constrained. Note that these assumptions do not
affect the qualitative nature of the results presented in this paper.

\subsection{Satellite galaxies}
\label{sec:satgal}
The satellite galaxies in the dark matter haloes are expected to inhabit
subhaloes. But resolving all the subhaloes is highly demanding in terms of
simulation resolution; therefore we will model satellite galaxies
following the dark matter distribution with little additional freedom. We
first assume that the number of satellite galaxies obeys Poisson
statistics. Current limited data sets do not allow us to test this
assumption, but it will be possible with future surveys. We assume that the mean
number of satellite galaxies depends only on the halo mass and is
independent of presence of satellites or centrals of other types of
galaxies in the neighbourhood. In the scenario when the cosmic web
influences satellites formation of a given type, one would expect both
of these assumptions of the fiducial model to be invalid. We will test if
data requires these assumptions to be changed by looking at
cross-correlations. In principle, a more complex model for the
satellite galaxies can be motivated based on current understanding of
galaxy formation and results from various hydrodynamical
simulations. But initially we will stay with this simple model; if it fails
to describe the observed data, then that will be evidence that a more complicated
satellite population model is needed. The mean number of satellite
galaxies as a function of halo mass is given by the following equation:
\begin{equation}
    \left< N_{\rm sat}^{\rm tot} \right>(M_{\rm halo})=\sum_{tr \epsilon TR} \left< N_{\rm sat}^{\rm tr} \right>(M_{\rm halo}),
\end{equation}
where the sum is over all different tracers in the sample. The number
of satellite galaxies per halo is given by following functional form:
\begin{equation}
    \left< N_{\rm sat} \right>(M_{\rm halo}) = \left( \frac{M_{\rm h} - \kappa M_c}{M_1}\right)^\alpha .
\end{equation}
The number of satellite galaxies is essentially assumed to be a power
law with index $\alpha$ and characteristic satellite mass $M_1$. The
parameter $\kappa$ sets a cut-off mass in units of $M_c$ below which
the probability of a satellite galaxy is zero. We use the same
functional form to model the number of satellites for all three tracers in
the sample, with independent parameters in each case. We do however require the existence
of a central galaxy in a halo before it is allowed to host any LRG satellite,
following \cite{zheng2005} and \cite{guo2015b}. But this assumption does not affect our results because LRG satellites inhabit massive halos with a central occupation probability of unity. The ELG and QSO satellite
galaxies can be hosted by haloes without any central galaxy, although this
rarely happens in practice. We also set $\kappa=1$ when ELG satellites are modelled with
the HMQ model based on the results from \cite{2018MNRAS.474.4024G}. One could
also imagine introducing correlations between satellites and centrals of
different tracers. We will consider some aspects of such correlations
later as an extension of the fiducial model.

The satellite galaxies are distributed following an NFW density profile for the
haloes, where the concentration for each halo is measured in the N-body
simulation (see Section ~\ref{sec:analysis}). We also assign isotropic
random velocities to satellites with the same velocity dispersion as the dark matter.
Due to the lack of small-scale clustering in the current data,
the results in this paper are insensitive to such choices and the data are not sufficient to inform us whether these assumptions need to be
modified.

Our approach is to use the simplest model unless observational data demand additional extension. Note that in our formalism we populate all galaxies simultaneously, so that it naturally avoids the non-physical situation of having central galaxies of different tracers in the same halo
at the same time. One of the ways we will test the validity of various
assumptions is by looking at the cross-correlations between different
tracers.

\section{Measurements and Systematics}
\label{sec:measurement}

\begin{figure}
    \centering
    \includegraphics[width=0.48\textwidth]{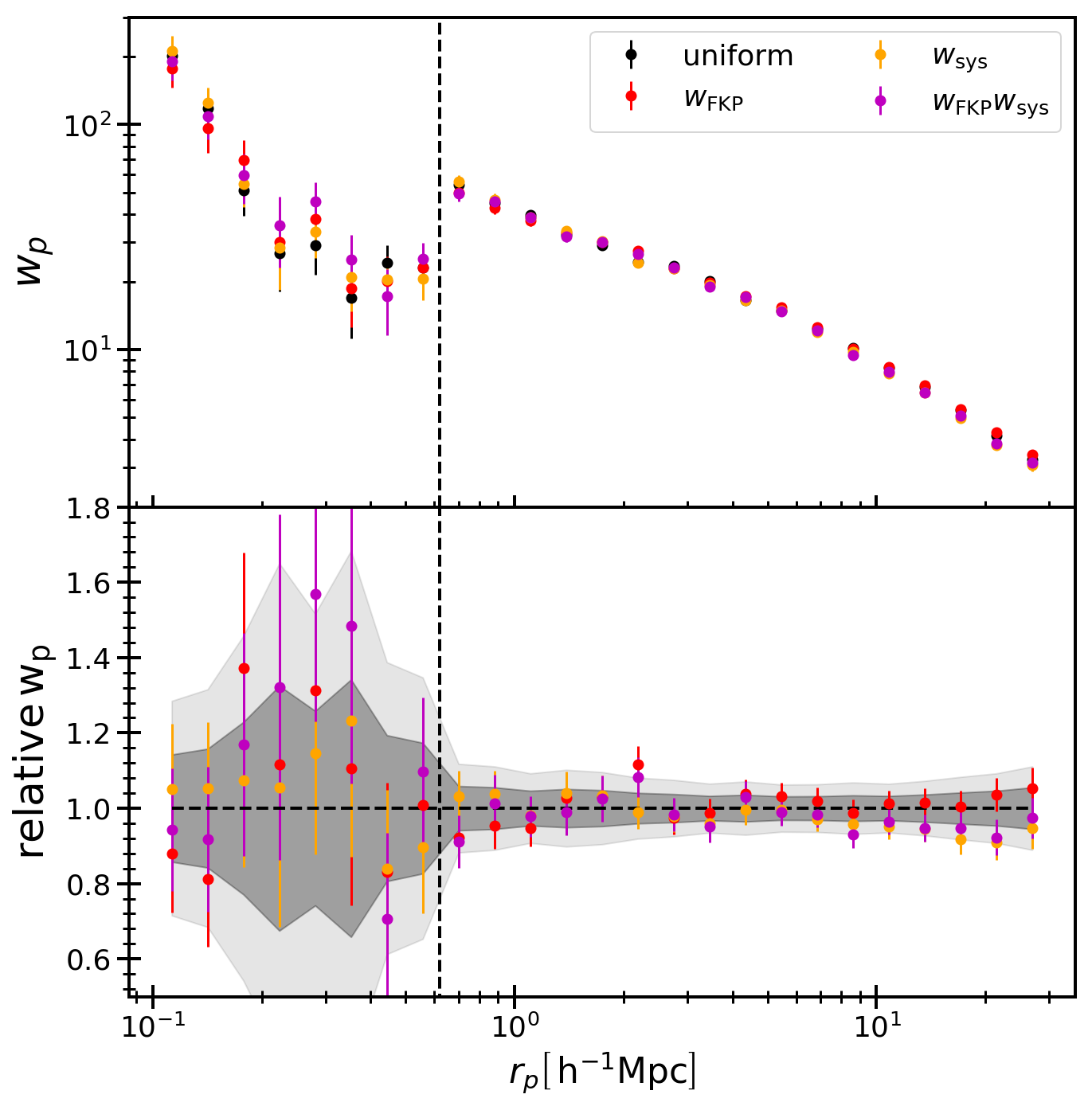}
    \caption{The projected correlation function for ELGs, using different
      systematic weights. This plot shows that the effect of the systematic
      weight is negligible in the $w_p$ measurements between scales of
      1--30$\mpcoh$. Therefore our results should be insensitive to the
      details of systematic weights of different samples. The
      black-dashed vertical line shows the fibre-collision scale for
      eBOSS at the mean redshift of 0.86.}
    \label{fig:wpsys}
\end{figure}

We first estimate the position of each galaxy in 3-dimensional space
by converting redshift to the line-of-sight distance using the
fiducial cosmology ($\Omega_m=0.307$, $h=0.67$).  We then measure galaxy
auto-correlation functions using the minimum variance Landay-Szalay
estimator \citep{LandySzalay93} given by:
\begin{equation}
    \xi_{\rm auto}(\vec{r})=\frac{DD(\vec{r})-2DR(\vec{r})+RR(\vec{r})}{RR(\vec{r})},
    \label{eqn:landay-szalay} 
\end{equation}
where $DD$, $DR$ and $RR$ are numbers of galaxy-galaxy, galaxy-random
and random-random pairs as a function of vector offset in 3-dimensional 
space. The cross-correlations are measured using following
estimator:
\begin{equation}
    \xi_{\rm cross}(\vec{r})=\frac{D_1D_2(\vec{r})}{D_1R_2(\vec{r})}-1, \label{eqn:xicross}
\end{equation}
where $D_1D_2$ and $D_1R_2$ are numbers of galaxy-galaxy counts and
galaxy-random counts. We note that redshift-space distortions make the
line-of-sight a special direction, and therefore we project the 3-dimensional 
space onto a 2-dimensional space that decomposes pair
separation vectors along the line-of-sight ($s_{\parallel}$) and
perpendicular to the line-of-sight ($s_{\perp}$). This gives us the 2-dimensional 
correlation function $\xi(s_\parallel,s_\perp)$. We then
measure the projected correlation function ($w_p$) by integrating the 2-dimensional
correlation function along the line-of-sight between
\smash{$s_\parallel=-40\mpcoh$} to \smash{$s_\parallel=+40\mpcoh$} and using 25
logarithmic bins in $s_\perp$ between $0.1\mpcoh$ and
$30\mpcoh$. The projected correlation function helps us constrain the
HOD parameters that govern the galaxy-halo connection. We create 86
jackknife regions for our sample and estimate jackknife covariances
wherever needed in this analysis.

The clustering measurement is sensitive to the completeness of the
observed galaxy sample for given selection. Therefore, it is important
to account for variations in the number of galaxies detected as a
function of position on the sky and as a function of various instrument level
patterns. The number of galaxies detected and the spectroscopic success
rate could both be correlated with stellar density, extinction, sky
brightness, air mass, position in the fibre plate etc. To remove these
correlations, \cite{AshleyRoss12} suggested the use of systematic
weights. We use a different combination of systematic weights which is
described in more detail in \cite{2018ApJ...863..110B} for LRGs and
\cite{2018MNRAS.477.1604G} for QSOs. Figure \ref{fig:wpsys} investigates if such systematics affect our measurements for ELGs. We find that the measurement of $w_p$
at small scales between 1--30$\mpcoh$ is insensitive to the details of
such systematic weights and hence the results described in this paper are not
prone to such systematics. We also note that fibre collisions lead to the
galaxy sample becoming highly incomplete below the fibre scale marked by the
vertical dashed line in Figure \ref{fig:wpsys}. There are several
methods for correcting the clustering measurements at these scales
\citep{2012MNRAS.427.3435A, 2012ApJ...756..127G,
  2017MNRAS.472.1106B}. For the purpose of this paper we will not use
these scales. We will perform this study in greater detail probing the
very small scales by applying the method developed by
\cite{2017MNRAS.472.1106B} in a future work.

We measure galaxy environments following the method described in
\cite{2019MNRAS.483.4501A}, which we briefly summarise here. For
galaxy survey one has to work with two catalogues. First a
catalogue of galaxies containing the 3-dimensional location of each
galaxy observed. Second, a catalogue of random positions representing
the volume observed by the survey with the density of random points at any
location representing survey completeness. In order to measure the galaxy
density we first count the number of random points in the Voronoi
tessellation of the galaxy field. We then estimate the local density as the inverse
of the random counts for each galaxy. This is then smoothed at the chosen
scale to determine the smoothed density. The density is then converted
to overdensity by first dividing by the mean density and then subtracting
1. In order to estimate the tidal environment of galaxies, we solve
the Poisson equation as described in Section 3.3 of
\cite{2019MNRAS.483.4501A} to obtain the tidal tensor,
whose eigenvalues are then measured. We then define the tidal anisotropy
($\alpha_5$) following \cite{2019MNRAS.483.4501A} as follows:
\begin{equation}
    \alpha_5=(1+\delta_5)^{-0.55} \sqrt{(\lambda_3-\lambda_2)^2+(\lambda_3-\lambda_1)^2+(\lambda_2-\lambda_1)^2}, 
    \label{eq:alpha5}
\end{equation}
where $\lambda_{1,2,3}$ are the eigenvalues of the tidal tensor field
and we adopt the convention that $\lambda_1<\lambda_2<\lambda_3$. The
quantity $\delta_5$ is the galaxy overdensity within a sphere of radius
$5\mpcoh$ centred on each galaxy. This form of tidal anisotropy was
first proposed in \cite{paranjape2017}, and it determines the level of
spherical anisotropy of the tidal field. The large value of $\alpha_5$
corresponds to tidally anisotropic regions, and small values correspond
to tidal isotropy. We always employ the same method to measure
environment in data and mock catalogues, so that our comparisons are
independent of any systematic uncertainty, for example the effects of
smoothing scale or peculiar velocity.

By looking at the combination of $\delta_8$ and $\alpha_5$ we can
assess the properties of galaxies in different parts of the cosmic web. For
example, clusters live in regions with high $\delta_8$ and
low $\alpha_5$, whereas voids will occupy regions with low $\delta_8$
and low $\alpha_5$. The regions with high $\alpha_5$ will be filaments
and we can probe the multi-scale nature of the cosmic web by looking at
high-$\alpha_5$ regions as the function of density. This will
correspond to filaments that have different densities. \cite{2019MNRAS.483.4501A} suggested to use the scale of $5\mpcoh$ for tidal anisotropy and $8\mpcoh$ for over-density and shown that results are not very sensitive to these choices.

\section{Obtaining the Best fit parameters}
\label{sec:analysis}

In this section, we describe the details of how we fit the model parameters and generate
the predictions for the auto- and cross-correlations.

We are using the publicly available MultiDark Planck (MDPL2) simulation
\cite{2012MNRAS.423.3018P} through the CosmoSim
database \footnote{\url{https://www.cosmosim.org/cms/simulations/mdpl2/}}.
MDPL is a dark matter only N-body simulation run using
the Adaptive-Refinement-Tree (ART) code \citep{1997ApJS..111...73K}.
The simulation assumes a flat $\Lambda$CDM cosmology with $\Omega_m=0.307$, $\Omega_b=0.048$, $h=0.67$, $n_s=0.96$ and $\sigma_8=0.82$. The calculation adopts a
periodic box of side 1000$\mpcoh$ and $3840^3$ particles. The
ART code used for MDPL is designed to preserve the physical resolution
to $~7\kpcoh$ for $z=0-8$. A halo catalogue using the
ROCKSTAR\footnote{\url{https://bitbucket.org/gfcstanford/rockstar}}
halo finder \citep{behroozi13} was constructed using the snapshot at
an effective redshift of $z = 0.86$ for MDPL2. ROCKSTAR starts with a
friends-of-friends group catalogue and analyses particles in full
phase space (i.e. position and velocity) in order to define halo
properties and robustly identify the substructures. We only use the main
haloes in the halo catalogue, removing all the subhaloes and modelling
satellite galaxies as described in Section \ref{sec:satgal}.

We then use the models described in Section \ref{sec:model} to predict
the number of central and satellite galaxies depending on the
mass for all haloes in the catalogue, which are then populated to
create a set of simulated catalogues, from which the
projected correlation
functions $w_p$ are measured. The $\chi^2$ fit
is then performed using the measured $w_p$ and jackknife covariance
matrix. We then run a minimizer algorithm to find the parameters which
give the minimum value of $\chi^2$ and hence the best-fit HOD
model. This process is repeated on only the auto-correlations of
the individual tracers and we do not use the cross-correlations in the
optimisation process. This is to test if the cross-correlation of
galaxies is any different from the prediction based on the halo model
estimated from the auto-correlations.

\section{Results}
\label{sec:result}

We fit the Multi-Tracer HOD (\mthod) model initially only to the auto-correlation of LRG, ELG
and QSO for our fiducial model, and we then compare the prediction of
cross-correlations from our model to the observed results. We then discuss
the signature of galactic conformity followed by the signature of the effect of the cosmic
web on quenching efficiency. The projected auto-correlations of LRGs,
QSOs and ELGs are fit between the scale of $1\mpcoh$ and $30\mpcoh$. The
best-fit model parameters for each tracer are shown in Table~\ref{tab:HODpar}.

\begin{figure*}
    \centering
    \includegraphics[width=1.0\textwidth]{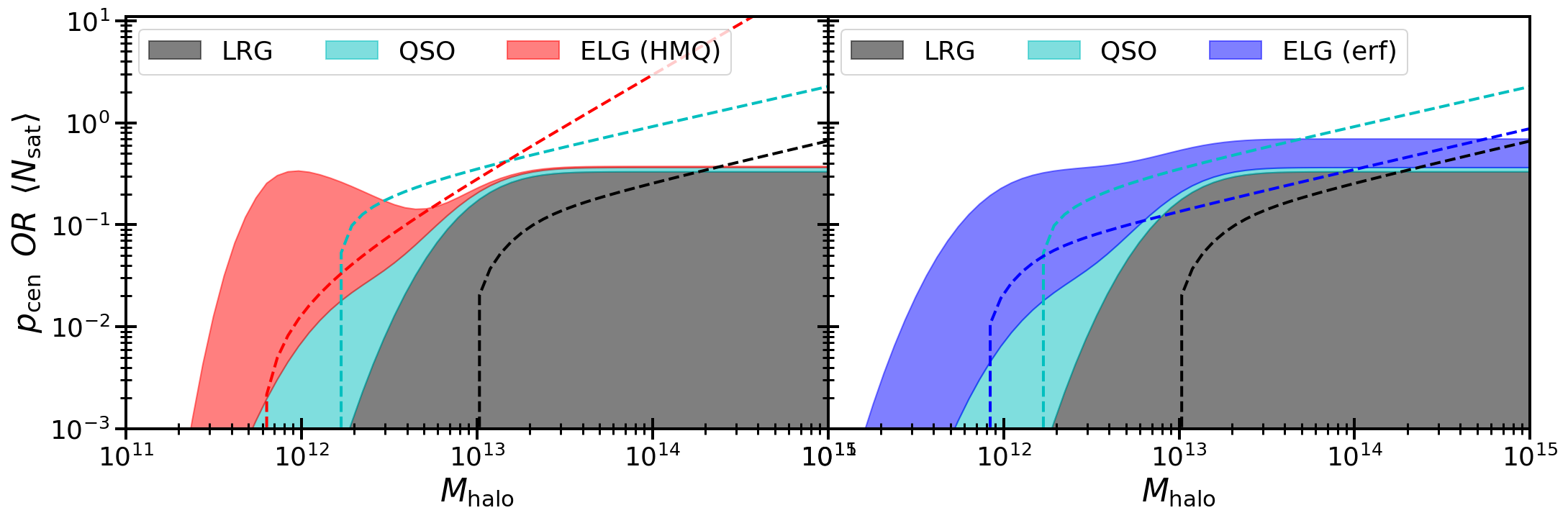}
    \caption{The best-fit occupation probability of galaxies as a
      function of halo mass in the \mthod\, framework. The left panel is when
      ELG centrals are quenched in massive haloes (HMQ model) and the
      right panel is for no quenching in massive haloes for ELG
      (the $\erf$ model). The shaded region is the occupation probability
      of centrals and the dashed line represents the mean number of satellite
      galaxies. The black, cyan and blue colours are for LRG, QSO and
      ELG respectively. This shows that LRGs inhabit massive
      haloes compared to QSOs, whereas ELGs live in less massive
      haloes compared to QSOs. }
    \label{fig:Hoccupation}
\end{figure*}

\begin{figure*}
    \centering
    \includegraphics[width=1.0\textwidth]{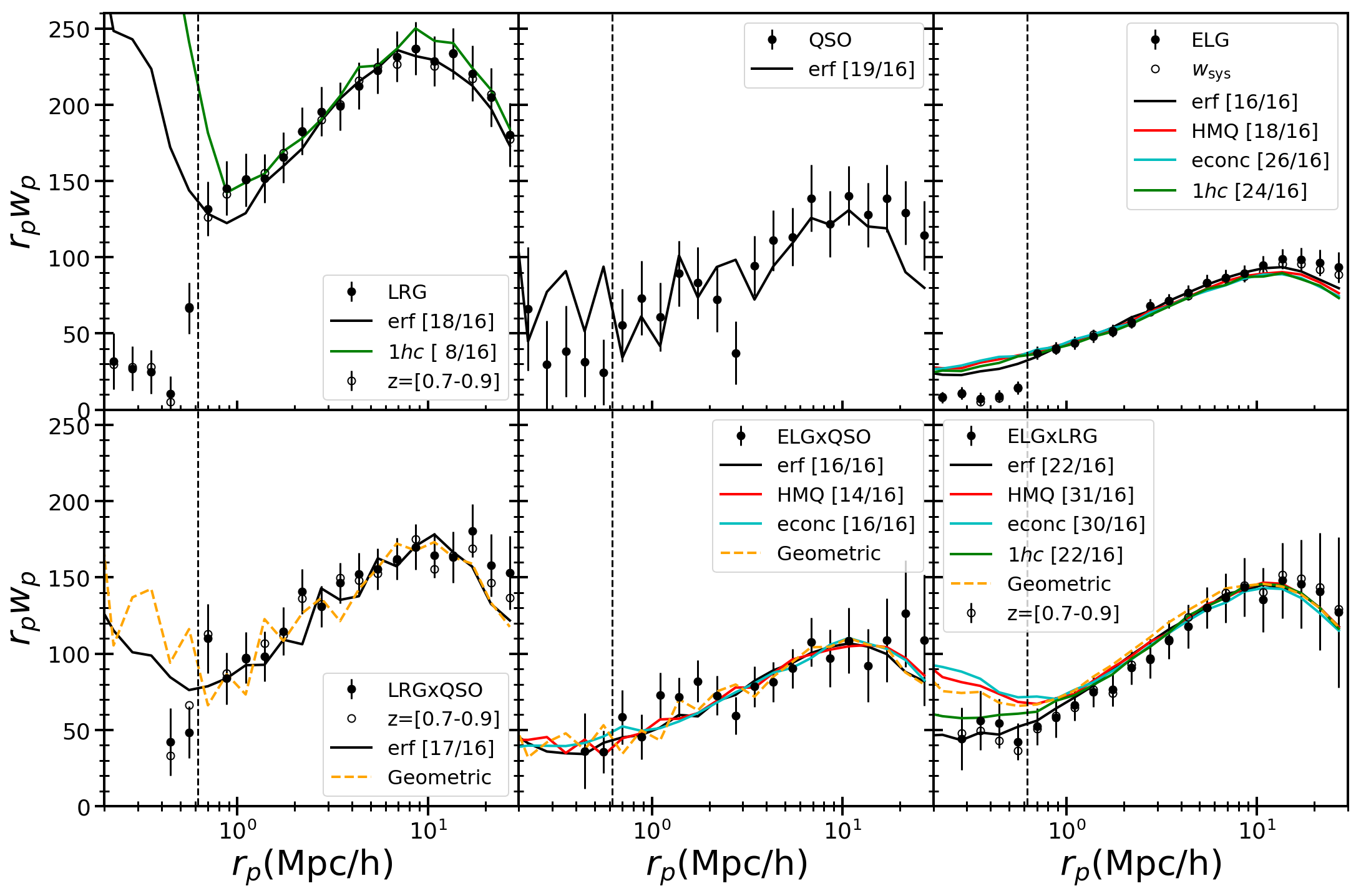}
    \caption{
    Projected correlation function $w_p(r_p)$ of auto- and
      cross-correlations for three tracers. The black points in each
      panel show the measurement from the eBOSS sample with jackknife
      errors. The black solid line shows the simple model where central galaxies in all
      three tracers are modelled by an error function. The red solid line
      shows the model in which ELGs are quenched in high mass
      haloes. The orange dashed line in the cross-correlation shows the
      geometric mean of auto-correlations. The green solid line shows
      a model with 1-halo conformity between ELG and LRG. The cyan solid line shows the model with stellite concentration for LRG (ELG) a factor of 2 higher (lower). The
      numbers in square brackets within the legend wherever shown are
      $[\chi^2/{\rm dof}]$. The black open circle in top-left, bottom-left and bottom-right are measurements from data with redshift cuts between 0.7 and 0.9. The black open circle in the top-right panel shows the ELG auto-correlation with systematic weights.}
    \label{fig:wp-auto-cross}
\end{figure*}

\begin{table}
	\centering
	\begin{tabular}{lcccc} 
		\hline
		Parameters & LRG & QSO & ELG ($\erf$) & ELG (HMQ)\\
		\hline
		$\log_{10}(M_c)$      & 13.0  & 12.21  & 11.88 & 11.75\\
		$\sigma_M$ & 0.60  & 0.60   & 0.56  & 0.58\\
		$\gamma$   & -     & -      & -     & 4.12\\
		$Q$        & -     & -      & -     & ${100}$\\
		$\log_{10}(M_1)$      & 14.24 & 14.09  & 13.94 & 13.53\\
		$\kappa$   & 0.98  & 1.0    & 1.02  &  ${1}$\\
		$\alpha$   & 0.40  & 0.39   & 0.40  & 1.0\\
        $p_{\rm max}$ & $\mathbf{0.33}$ & $\mathbf{0.033}$ & $\mathbf{0.33}$ & $\mathbf{0.33}$ \\
		\hline
	\end{tabular}
	\caption{The best fit \mthod\, parameters for different tracers. The last column shows the ELG with HMQ model in which ELG galaxies are quenched at high mass haloes.}
	\label{tab:HODpar}
\end{table}

We first show the results of measurements and best-fit models in
Figures~\ref{fig:Hoccupation} and~\ref{fig:wp-auto-cross}. 
We show the halo occupation statistics of all three tracers for
centrals and satellites in Figure~\ref{fig:Hoccupation}. The plot on
the left shows the occupation when ELGs are populated using the HMQ model and
the one on the right shows the behaviour when ELGs are treated using the $\erf$ model. The shaded region
represents the occupation probability of central galaxies, with grey area standing for
LRGs, cyan for QSOs, blue for ELGs  with $\erf$ model and red for ELGs with HMQ model. 
The dashed line shows the mean satellite occupation as a function of halo mass.
This shows that the LRGs inhabit massive haloes compared with QSOs, whereas ELGs
live in less massive haloes than QSOs. 
We also note that when ELG centrals are quenched in the massive haloes (using the HMQ model) then the mean occupation of satellites for ELGs in massive haloes increases (by comparing red dashed line in the left panel with the blue dashed line in the right panel of Figure~\ref{fig:Hoccupation}). 
This suggests that quenching could trigger outflows of cold gas from halo centres, leading to increased star formation in the galaxies that are on the outskirts of massive haloes. Given the occupation probability we also
note that it is very unlikely to have an ELG central and LRG satellite
in the HMQ model because the distributions in halo mass are almost disjoint. Another interesting point to note is that in the left panel the total probability of a halo hosting a central galaxies has a peak at around $6 \times 10^{11} \msolaroh$ and then a plateau for haloes with mass above $10^{13} \msolaroh$. This implies that the total efficiency of galaxy formation for star forming eBOSS ELGs and eBOSS LRGs is suppressed around $3\times 10^{12} \msolaroh$. This probably arises from a combination of eBOSS selection and  green valley galaxies dominating this intermediate halo mass.

The projected auto- and cross-correlations are shown in
Figure~\ref{fig:wp-auto-cross}. The top panels from left to right are
for LRG, QSO and ELG auto-correlations. The bottom panels are for
LRG$\times$QSO, ELG$\times$QSO and ELG$\times$LRG cross-correlations from left to
right. The $x$-axis is the projected separation and the $y$-axis is the
projected correlation function multiplied by projected separation for
clarity. The vertical dashed line in each panel indicates the fibre
collision scale below which the clustering measurements are biased due
to incomplete spectroscopic sample, and hence not used in this study. The
black points are the observed eBOSS data with error bars estimated from
the jackknife sub-sampling method. The black solid line shows the simple
model where central galaxies of all three tracers are modelled by
the $\erf$ function. The red solid line shows the model in which ELGs are
quenched in high mass haloes. The orange dashed line in the
cross-correlation plots shows the geometric mean of the auto-correlations. The
numbers in square brackets within the legend wherever shown are
$[\chi^2/{\rm dof}]$ of the model. Both models provide an excellent
fit to the auto-correlation.
We can notice from the auto-correlation plot that the clustering
amplitude of LRGs is highest, then QSOs followed by ELGs. This means
that the bias of LRGs QSOs and ELGs will be ordered in declining order and hence that the mean
halo mass will have a trend in the same direction. This is consistent with the picture from Figure~\ref{fig:Hoccupation}, showing mean halo mass of ELGs (LRGs) is lower (higher) than mean halo mass of QSOs.
We also note that the error on the ELG auto-correlation
is smaller than for LRGs, with QSOs showing the largest error; this is because
the number density of these tracers follows that order. The
cross-correlation predicted by the best-fit model is consistent with
observations. 
This means that no imprint of complicated formation mechanism of QSOs
is detected in the eBOSS sample, given that the halo model predicts the
cross-correlation of QSOs with both LRGs and ELGs at all measured
scales. The cross-correlation predicted for ELG$\times$LRG by our best-fit HMQ model shows a $\chi^2/{\rm dof}=31/16$ and hence is not a good
fit. This indicates that if the quenching in ELGs is driven by halo mass (as assumed in the HMQ model), then the occupation of LRGs and ELGs cannot be independent as assumed by the fiducial model. This is probably a
signature of galactic conformity. The green line shows the model with
1-halo galactic conformity which improves the goodness of fit and is
discussed in Section~\ref{sec:galactic-conformity}. Such an effect can possibly also arise by the distribution of satellite galaxies being different from NFW. We study a model in which LRG satellites has a factor of 2 higher concentration and ELG satellites  a factor of 2 lower concentration shown by cyan solid line in Figure~\ref{fig:wp-auto-cross}. We found that such a change in the model has no effect in the predicted signal and hence cannot explain the cross-correlation between LRGs and ELGs.

The number density of LRGs drops sharply above redshift of 0.9 as shown in Figure~\ref{fig:coverage}. Therefore, to understand if this has any effect on our measurement we repeated our measurements of LRG auto-correlation and its cross-correlation with ELGs and QSOs by selecting all galaxies within redshift range of 0.7 and 0.9. The black open circles in the top left, bottom left and bottom right panels show the LRG auto-, LRG$\times$QSO and ELG$\times$LRG cross-correlations  for this sub-sample respectively. This shows that restricted redshift selection and hence the sharp drop in LRG number-density at higher redshift does not affect our measurements. The systematic weights in ELG sample affect large-scale clustering significantly. Therefore, we also show the ELG auto-correlation with systematic weights by black open circles in the top right panel. This confirms that for the purpose our study in this paper the impact of systematic correlation of ELG number density with sky conditions is negligible.

\subsection{Galaxy properties from \mthod}
Based on the \mthod\, model we find that the luminous red galaxies in the
eBOSS sample live in especially massive haloes, with $1\%$ of the galaxies
living in haloes with mass below $2.2 \times 10^{12} \msolaroh$.
Haloes of mass $2.8 \times 10^{15} \msolaroh$ host on average one LRG satellite
galaxy. The mean halo mass of the LRG sample is $1.9 \times 10^{13}
\msolaroh$.  The LRG sample has a satellite fraction of $17\%$.
\cite{2017ApJ...848...76Z} also studied an earlier version of the eBOSS LRG
sample and found the mean halo mass of the sample to be $2.5 \times 10^{13}
\msolaroh$ with a satellite fraction of $13 \pm 3\%$. The slightly
higher mean halo mass and lower satellite fraction in
\cite{2017ApJ...848...76Z} compared to our study can plausibly be understood as being due
to the difference between our redshift cut of 0.7 compared to 0.6 used in the
earlier study. \cite{2017ApJ...848...76Z} also showed that despite the
possible incompleteness in the LRG sample, the HOD approach is sufficient
to analyse this sample.

The QSOs in the eBOSS sample have lower bias than the LRGs and hence extend to
lower halo masses than LRGs. The mean halo mass of the QSO sample is $5
\times 10^{12} \msolaroh$, and $99\%$ of QSOs in our sample are found
in haloes with mass above $4.2 \times 10^{11} \msolaroh$. The
characteristic halo mass of satellite QSOs is $1.3 \times 10^{14}
\msolaroh$, for which the mean satellite number is 1. The QSO satellite
fraction is $34\%$. Typically the form of the QSO HOD model is uncertain,
but we expect our results not to be very sensitive to the parametric
form of the HOD. The mean halo mass of the eBOSS QSO sample reported in
\cite{2017MNRAS.468..728R} is $5 \times 10^{12} \msolaroh$, which is
in agreement with our measurement despite a very different model. But
our best-fit satellite fraction is much higher than generally reported
in other QSO studies. We discuss the implication of such high QSO
fractions and obtain more robust constraints in our companion paper
(Alam et. al. in prep.).

The ELGs in the eBOSS sample are star-forming galaxies. We expect the
galaxy quenching mechanism to be more effective in higher-mass haloes, so
it is less likely for an ELG galaxy to be a central galaxy. We
have used two different models for ELGs: the first one (`$\erf$')
does not have any effective quenching, but the second one (`HMQ')
assumes galaxies to be quenched at high halo masses. Both models provide
good fits to the auto-correlation of the ELG sample. The minimum halo mass
required to cover $99\%$ of ELG galaxies is $2.1 \times 10^{11}
\msolaroh$ for the $\erf$ model and $3.2 \times 10^{11} \msolaroh$ for
the HMQ model. The mean halo mass of the ELG sample is $2.9 \times 10^{12}
\msolaroh$ for the $\erf$ model and $1.1 \times 10^{12} \msolaroh$ for
the HMQ model. The characteristic satellite halo masses are $1.4
\times 10^{15} \msolaroh$ for $\erf$ and $3.6 \times 10^{13}
\msolaroh$ for HMQ. The satellite fraction of the ELG sample is
$12\%$ and $17\%$ for the $\erf$ and HMQ models, respectively. One
interesting point this raises concerns the minimum halo mass that one would
require in order to host an ELG. In the HMQ model the mean halo mass of an ELG is 3
times lower than in the $\erf$ model. This means for future surveys like
DESI the simulation requirement could have significant impact on the
resolution requirement, given that the minimum host halo mass of $99\%$ of
the ELG sample could be higher by a factor of 2 for the more physical HMQ model. 
\cite{2019ApJ...871..147G} report that the mean halo mass of ELGs is $10^{12} \msolaroh$ with a satellite fraction between $13\%$ to $17\%$, consistent with the results obtained in this paper.

\subsection{Galactic conformity}
\label{sec:galactic-conformity}


Galactic conformity is usually identified in observations by studying
the colours or star formation rates of galaxies, and quantified in
terms of red (quiescent) fractions and/or quenching efficiencies.
In testing the concept of galactic conformity, it is crucial to compare samples
that have been matched in one or more parameters in order to avoid
trivially arising correlations.  Following the probabilistic
description of the meaning of galactic conformity, put forward by
\cite{Knobel2015}, this phenomenon can be seen as arising due to one
or more \textsl{hidden variables} not accounted for in the analysis.
Among obvious candidates for these potential hidden parameters that
have been addressed in previous studies are the luminosity, stellar
mass, halo mass of galaxies or their local density. None of them has
been identified as a principal driver of the observed phenomenon.

Numerical simulations allow us to extend the range of explored
parameters to properties that are not easily accessible in the
observed data sets, such as age or formation history of galaxies and
their haloes.  Using the \illustris \citep{vogelsbergeretal14} simulation,
\cite{Bray2016MNRAS.455..185B} confirmed a significant signal of
galactic and halo conformity out to distances of about 10~Mpc that
coupled with a galaxy colour-halo age relation, resulting in the reddest
galaxies preferentially residing in the oldest haloes. The
interpretation of galactic conformity in this picture is that dark
matter clustering is the primary factor, but a sufficiently tight
galaxy colour-halo age relation is necessary in order to measure the
conformity signal.  Similarly, \cite{Mika2018} found that the
cosmological hydrodynamical simulation \mufasa\citep{Dave2016} produces conformity in
various galaxy properties at fixed halo mass, with the \textsl{1-halo} term
dominating the signal.

Commonly discussed interpretations of galactic conformity seen in the
data, on large scales in particular, invoke assembly bias leading
to the dependence of galaxy properties on the properties of a halo beyond its
mass.  Among these, halo age and concentration correlated with galaxy
colour or star formation history were found to reproduce the galactic
conformity in analytical models based on the HOD framework
\citep[e.g.][]{Hearin2015,Paranjape2015,Pahwa2017}.  On the other
hand, using the HOD halo quenching framework of \cite{zu2015,zu2016},
\cite{Zu2018} showed that the galaxy conformity seen in the SDSS data can
be naturally explained by the combination of halo quenching and the
variation of the halo mass function with environment, without the need for any
galaxy assembly bias.  This is consistent with the findings of
\cite{Tinker2018}, who compared the conformity in the SDSS with the
predictions from the halo age-matching model and concluded that it can be
produced by mechanisms other than halo assembly bias, e.g. difference
in halo mass at fixed stellar mass for blue (star-forming) and red
(passive) galaxies. Such a bimodality is suggested by the weak lensing
measurements of bright galaxies in the SDSS, where at fixed stellar
mass, red centrals are found to preferentially reside in more massive
haloes than blue ones \citep{zu2016}. Applying such a colour dependent
halo-to-stellar-mass ratio \citep[e.g.][]{zu2015} to estimate the halo
mass of galaxies in the GAMA survey, \cite{Treyer2018} recently showed
that conformity at fixed halo mass indeed vanishes.

Galactic conformity will cause the cross-correlation of LRGs and ELGs to
differ from the behaviour expected in a universe without such conformity. We detect a signature
of a failure of our fiducial model (which lacks conformity effects) to describe
the cross-correlation of ELG and LRG samples. Therefore we introduce an explicit \textsl{1-halo} and \textsl{2-halo} galactic conformity in our model to constrain this effect. One of the major systematics for conformity studies in
the past was uncertainty in the identification of central and
satellite galaxies. We avoid such issues by introducing conformity in
our model only where we precisely know which galaxies are centrals and satellites,
and comparing only the overall auto- and cross-correlations.

\subsubsection{\textsl{1-halo} galactic conformity}
\begin{figure}
    \centering
    \includegraphics[width=0.48\textwidth]{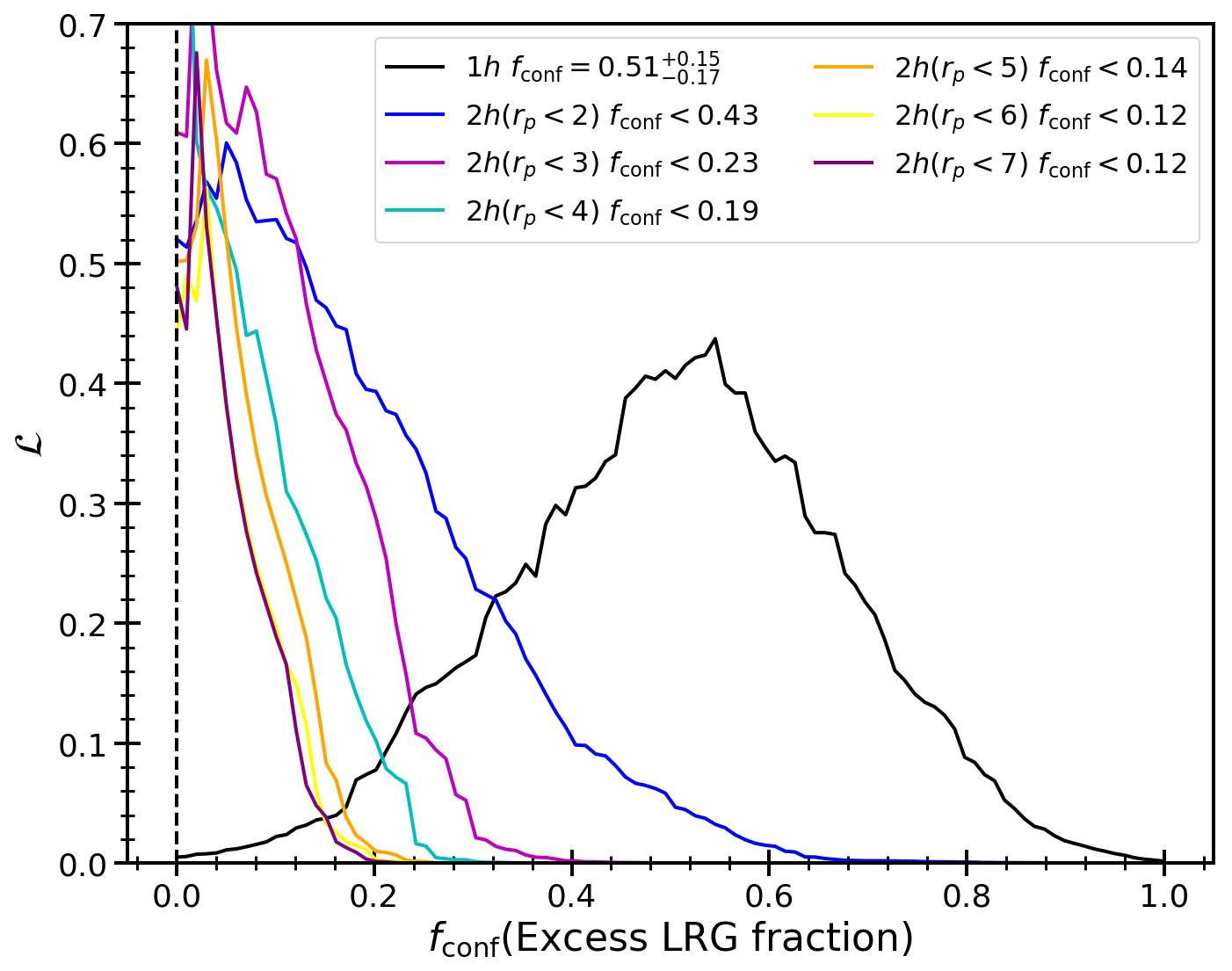}
    \caption{Likelihood of the galactic conformity parameter $f_{\rm conf}$. The black dashed line represents the zero conformity model. The black solid line shows the \textsl{1-halo} conformity constraints and other coloured solid lines are for \textsl{2-halo} conformity at different scales as shown in the legend. We also provide $68\%$ constraint for \textsl{1-halo} conformity and $95\%$ upper limit for \textsl{2-halo} conformity in the legend.}
    \label{fig:conf_like}
\end{figure}

We therefore introduce an additional parameter in our model denoted by $f_{\rm conf}$,
which represents the level of galactic conformity. In this model
we say that in all haloes with an LRG as a central galaxy, a certain
fraction of ELG satellites will be turned into LRGs. Mathematically this can be written as follows for haloes with an LRG as the central galaxy:
\begin{align}
 \left< N_{\rm sat}^{\rm LRG, conf} \right> &=   \left< N_{\rm sat}^{\rm LRG} \right> +
 f_{\rm conf}\left< N_{\rm sat}^{\rm ELG} \right>, \\
 \left< N_{\rm sat}^{\rm ELG, conf} \right> &= (1-f_{\rm conf})\left< N_{\rm sat}^{\rm ELG} \right>.
    \label{eq:gconf}
\end{align}
We do not enforce a number density constraint here due to the large uncertainty in the number density; the impact of this change is small compared to the uncertainty. One can also argue using a more complicated model of conformity. But here we limit ourselves to this simple model due to lack of any strong motivation for something more complicated. 
We first look at all haloes with LRG centrals and count the total number of ELG satellites in such haloes. We then randomly convert a fraction $f_{\rm conf}$ of those ELG satellites into LRGs. This model is then fitted to
the projected correlation function of LRG, ELG and ELG$\times$LRG, thus determining
the relative likelihood as a function of $f_{\rm conf}$. Figure~\ref{fig:conf_like} shows the
likelihood of $f_{\rm conf}$ for the \textsl{1-halo} conformity signal with a black
solid line. We measure $f_{\rm conf}= 0.51^{+0.15}_{-0.17}$, a
3$\sigma$ detection of \textsl{1-halo} galactic conformity at a mean redshift of
0.86. One of the consequences of \textsl{1-halo} conformity is that it changes the satellite fraction of the sample. The best fit model with quenching shows satellite fraction of 20.7\% and 16.6\% for LRGs and ELGs respectively. Whereas the model without conformity showed satellite fraction of 17\% for both LRGs and ELGs.

The \textsl{1-halo} conformity is typically compared at fixed halo mass \citep[e.g.][]{Treyer2018}. In this work we do not control for the halo mass due to difficulties in obtaining such measurements. The conformity model introduced in this work allows for galactic conformity only in haloes with central galaxies that are LRGs, which have a mean halo mass of $1.9 \times 10^{13} \msolaroh$. Therefore, the halo mass for \textsl{1-halo} conformity is being fixed indirectly using the mean mass of LRGs.

\subsubsection{\textsl{2-halo} galactic conformity}
We now use the same parameter $f_{\rm conf}$ to model \textsl{2-halo}
galactic conformity. In this case we again look at all LRG centrals
and count the number of ELGs (both centrals and satellites) within a given
distance of the central LRG galaxies. The model then demands that a fraction
$f_{\rm conf}$ of those ELGs should be turned randomly into LRGs.
This modified galaxy catalogue is then used to measure the
clustering, which is fitted to constrain $f_{\rm conf}$. Figure
\ref{fig:conf_like} shows the likelihood of $f_{\rm conf}$ for
\textsl{2-halo} conformity as a function of scale. We do not detect any
\textsl{2-halo} galactic conformity signal, and only obtain an upper limit on
the possible size of the effect.
The legend in the plot quotes $95\%$ upper
limit on the \textsl{2-halo} galactic conformity parameter.

\subsection{Galaxy quenching and environment}
\begin{figure}
    \centering
    \includegraphics[width=0.48\textwidth]{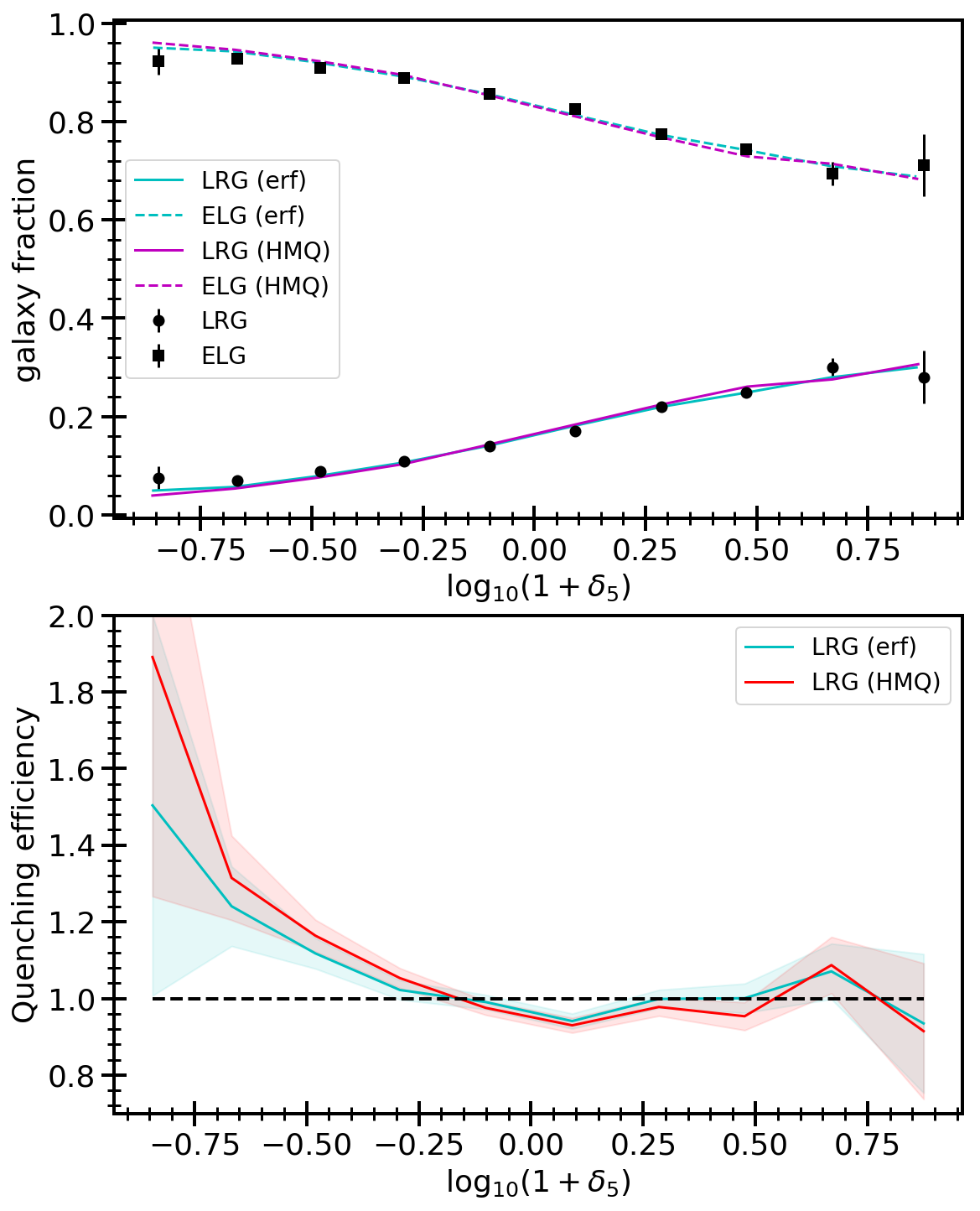}
    \caption{The environmental dependence of galaxy quenching. The top panel shows the fraction of ELGs and LRGs as a function of environmental overdensity for the observed data with black points, and model predictions with coloured lines. The bottom panel shows the excess quenching efficiency with observed overdensity compared to models.}
    \label{fig:delta-gfrac}
\end{figure}

\begin{figure*}
    \centering
    \includegraphics[width=1.0\textwidth]{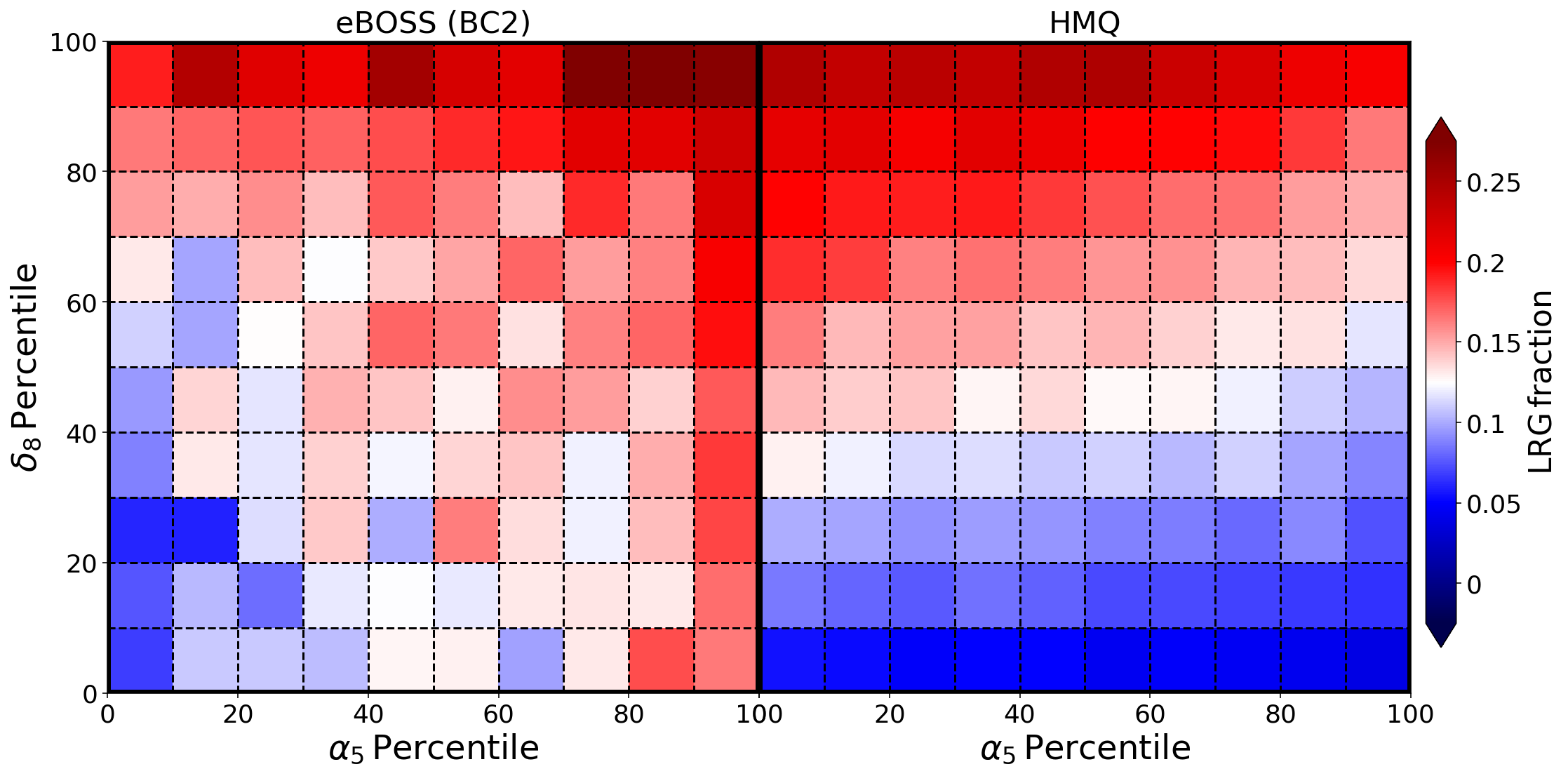}
    \caption{Comparison of the LRG fraction between eBOSS (left) and
      HMQ (right) galaxies in the $\delta_8-\alpha_5$ plane. Each cell
      is colour-coded by the fraction of LRGs compared to all galaxies
      in 2D bins of ten percentiles in $\delta_8$ and $\alpha_5$
      jointly. The subtle dependence of the LRG fraction on $\delta_8$
      mostly reflects halo mass driven quenching as it is
      reproduced in the HMQ results,  but the dependence on $\alpha_5$ is probably the
      evidence for a tidal dependence of galaxy quenching. The typical
      errors in the LRG fractions are 0.02 for data and 0.005 for HMQ
      model. }
    \label{fig:delta-alpha-gfrac}
\end{figure*}

The environmental dependence of galaxy properties is one of the key aspects of
the galaxy formation process. In general the properties of galaxies
strongly correlate with their host halo mass and overdensity,
reflecting the variation of the gravitational potential with their location.
As the major force during galaxy
formation is gravity, the local dark matter density or the host dark matter halo of galaxies are expected to drive their properties, in particular governing the extent of quenching.
However, most hydrodynamical simulations predict that at
some point baryonic physics should play a role, so that such processes need to be
accounted for in addition to the dark matter halo mass in order to understand galaxy
quenching. But given that such processes are second order, the net effect
is often small and hard to disentangle from the dependence on the dark
matter only \citep{2019MNRAS.483.4501A}. 

Therefore we make a detailed comparison between the observed
eBOSS sample and the \mthod\, model, the simple form of which assumes quenching
to be driven by halo mass alone.
First we measure the LRG fraction as a function of
density contrast ($\delta_8$) in data and in the two mocks using the
method described in Section \ref{sec:measurement}. We show LRG and ELG
fractions with overdensity in the top panel of Figure
\ref{fig:delta-gfrac} with black points for the eBOSS sample and with
coloured solid lines for different models. If we think of the fraction of
LRGs in a given environment as an indicator of quenching efficiency, then
from this plot we first show that the galaxy quenching efficiency
strongly depends on the dark matter overdensity -- a trend that is also
reproduced by the \mthod\, catalogue as shown with the coloured lines. The
fact that the \mthod\, catalogue shows a reasonable agreement with the data means
that the dominant terms causing this dependence must arise from the dark
matter halo mass. We note that in high-density environments the data
and models agree very closely, meaning that
galaxy quenching is mainly driven by halo mass in these regions of high density. However, in
low density environments the data exhibit more efficient quenching than the
model predicts. This is highlighted further in the bottom panel of
Figure \ref{fig:delta-gfrac}, showing the ratio of the observed LRG fraction
with respect to the model. The bottom panel shows that the galaxy
quenching efficiency is higher in the data by up to a factor of 1.5 for the
more physical HMQ model, and by a factor of 1.2 in the $\erf$ model. In
addition, the quenching efficiency around the mean density is lower in
the data than in the mocks. This is possibly caused by the fact that
the material in under-dense regions will deplete faster than in higher
density regions due to outflows; thus the quenching efficiency is
enhanced. This also results in slightly less efficient quenching in
the mean density regions due to the transfer of extra gas from low
density regions to the mean density regions.

We further ask more specifically whether quenching of galaxies is influenced by
{\it geometrical\/} location within the cosmic web. The accretion of gas onto haloes can depend on the
tidal environment of the haloes. This can cause a further dependence of quenching efficiency 
on tidal environment, in addition to the effect of halo mass
\citep{keres2005}. To investigate whether any cosmic web quenching
exists, we look at the fraction of LRGs in different cosmic web
environments, defined by $\delta_8$ and $\alpha_5$. We first evaluate
$\delta_8$ and $\alpha_5$ values for both data and mocks, following the
method described in section~\ref{sec:measurement}. We then split the
galaxies into 10 percentile bins of $\delta_8$ and $\alpha_5$
jointly. For each such cell, the fraction of LRGs is estimated by taking the
ratio of the number of LRGs to the number of all galaxies in the given
cell. This 10$\times$10 matrix of LRG fractions is shown in
Figure~\ref{fig:delta-alpha-gfrac}, where the $x$-axis is for the $\alpha_5$
percentile and the $y$-axis is for the $\delta_8$ percentile. The left panel shows
the 2D LRG fraction measured from eBOSS data and the right panel
shows the corresponding result derived
from the HMQ mocks. Note that the tidal anisotropy ($\alpha_5$) is by
construction independent of $\delta_8$ (see Figure~\ref{fig:bc-hist2d} for details). 
It is clear that the observed LRG
fraction depends on both $\delta_8$ and $\alpha_5$ simultaneously,
showing several interesting features in the $\delta_8$ vs. $\alpha_5$
plane:
\begin{itemize}
    \item The high-density filaments (high $\delta_8$ and
      high $\alpha_5$) and clusters (high $\delta_8$ and
      low $\alpha_5$) show strong quenching efficiency in both data and
      model. But the data show slightly lower quenching efficiencies in
      clusters compared to high-density filaments, whereas the model
      shows slightly higher quenching efficiencies in clusters compared
      to the data. This implies that the quenching efficiency in high-density
      regions is largely driven by halo mass, with additional small effects
      from the cosmic web in the filaments.
    \item The observed quenching efficiency in void regions
      (low $\delta_8$ and low $\alpha_5$) is small and is similar to the
      quenching efficiency in the mock. Therefore, the quenching
      efficiency in voids can be explained by halo mass driven
      quenching.
    \item The observed quenching efficiency in filaments
      (high $\alpha_5$) is high and shows a weak positive correlation
      with density ($\delta_8$). This is in contrast with the mock
      assuming halo mass driven quenching, which predicts the quenching
      efficiency in filaments to correlate strongly with density
      ($\delta_8$). This suggests that the cosmic web has a significant
      impact on the quenching efficiency in low-density filaments.
    \item Galaxies in mean density isotropic regions (intermediate $\delta_8$
      and low $\alpha_5$) show lower quenching efficiency in the data
      compared to the model predictions.
\end{itemize}

The halo mass driven quenching produces a quenching efficiency map
with a complicated dependence on cosmic web properties, as shown in the right panel of
Figure~\ref{fig:delta-alpha-gfrac}. This matches the observed
quenching efficiency map (shown in the left panel of
Figure~\ref{fig:delta-alpha-gfrac}) over most of the parameter plane. But it
under-predicts the quenching efficiency in the low-density filaments. This
implies that either the galaxy quenching efficiency for-low density
filaments is driven by mechanisms beyond halo mass quenching, or possibly that the star forming galaxies in low-density filaments occupy different part of colour-magnitude space compared to other star-forming galaxies. Either way this is an interesting signature, showing the impact of the cosmic web on
galaxy formation and evolution.

\section{Summary and Discussion}
We have extended the halo occupation distribution  model into a form suitable for application to multiple tracers in ongoing and
future surveys: the Multi-Tracer HOD (\mthod). The model by
default includes the environmental dependence of the halo mass function
by using numerical halo catalogues from N-body
simulations to construct mock galaxy data sets. We propose a new parametric form for the occupation
probability of star-forming galaxies, which allows the incorporation of quenching physics: the HMQ model
(see equation~\ref{eq:NHMQ}). We apply this \mthod\, framework to the eBOSS data
within the redshift range 0.7--1.1. We first obtain halo model
constraints for three kinds of tracers in the eBOSS sample: LRGs, ELGs and QSOs.
We compare our results with earlier efforts to model these
samples, finding generally good agreement. The mean halo masses of
LRGs and QSOs are $1.9\times 10^{13}\msolaroh$ and $5\times
10^{12}\msolaroh$ respectively, in agreement with previous studies
\citep{2017ApJ...848...76Z,2017MNRAS.468..728R}. Within the HMQ model, the mean halo
mass of ELGs is found to be $1.1\times 10^{12}\msolaroh$. We
also note that in the absence of any quenching at the centres of massive haloes
the mean halo mass is a factor of 3 larger for ELGs.

We then focus on the cross-correlations between different tracers,
comparing our model with the data. The overall cross-correlations are
in very good agreement at large scales. The
cross-correlation of QSOs with ELGs (star-forming) and LRGs (quenched)
is completely consistent with the prediction from the \mthod\,
models. We see no signature of the eBOSS QSOs being especially
located around star-forming galaxies within the statistical error in
the data. This supports the arguments that QSOs are formed in typical galaxies.

However, we found that the cross-correlation between LRGs and ELGs at small scales
is not in such good agreement with our fiducial model. The deviation between the
observed cross-correlation and prediction is limited to scales below $5 \mpcoh$,
which is similar to the signature of galactic conformity. We hypothesise
that this deviation reflects the presence of galactic conformity in the
observed sample. We introduce galactic conformity within the \mthod\,
framework as described in Section~\ref{sec:galactic-conformity}. We
compared the resulting predicted cross-correlation with the data,
obtaining consistent results (see the green line in
Figure~\ref{fig:wp-auto-cross}). This indicates the detection of 
\textsl{1-halo} galactic conformity at the $3\sigma$ level, although no signature of a conformity
effect in the \textsl{2-halo} regime was detected (see
Figure~\ref{fig:conf_like}). We also looked at the possibility that
the deviation in the cross-correlation might arise simply because the
concentration of LRG and ELG satellites are respectively smaller and
larger than the concentration of the host dark matter haloes. We created a
mock catalogue with double the satellite concentration for ELGs and
half the satellite concentration for LRGs. The resulting
cross-correlation is not different at the relevant scale and could not
explain the galactic conformity signal. Thus
the \textsl{1-halo} galactic conformity signal we observe cannot
be the result of differences in the satellite galaxy distributions within dark
matter haloes.

We then study the environmental dependence of quenching by comparing the
fraction of LRGs as a function of the cosmic web environment between the
observed data and the \mthod\, mock galaxy catalogue. The cosmic web is
characterised by galaxy over-density ($\delta_8$) and tidal anisotropy
($\alpha_5$) at the position of each galaxy. The estimate of the tidal
anisotropy ($\alpha_5$) for galaxies around the survey boundary may become
unreliable. We show the impact of boundary effects in
Appendix~\ref{sec:bc-effect} and remove galaxies around survey boundaries
in order to avoid any bias in our results. We show that in the high density
regime the halo mass driven quenching in the \mthod\, model is consistent with
observations. But in lower density regions we find that the observed
quenching efficiency shows a deviation from halo mass driven
quenching. We found that the quenching efficiency in the \mthod\, model is
smaller by a factor of 1.5 compared to the observed data in under-dense regions (see Figure~\ref{fig:delta-gfrac}). 

To understand
if the difference in quenching efficiency arises from any specific
tidal environment, we study the LRG fraction in the two-dimensional space
of density and tidal anisotropy as shown in
Figure~\ref{fig:delta-alpha-gfrac}. We found that the observed galaxy quenching
efficiency depends on both overdensity as well as
tidal anisotropy. The quenching efficiency in clusters, voids and high
density filaments is consistent between observed data and our mock
catalogue. This suggests that in such parts of cosmic web, galaxy
quenching is well explained by the halo mass driven quenching model. But
the quenching efficiency in the low-density filaments is predicted to
be much smaller than the observed effect. This suggest that either the
galaxy quenching efficiency is driven by tidal fields beyond halo mass
in the low-density filaments, or that the OII-bright star-forming galaxies (ELGs) observed by eBOSS avoid
low-density filaments due to the impact of cosmic web on galaxy
formation. In any case this is a clear signature of tidal fields playing a role
in the galaxy formation process and dominating in the
low-density filaments. The role of the filaments, beyond that of density,
was recently highlighted by works showing that red or high-mass galaxies tend to be closer to the filaments than blue or low-mass and late-forming galaxies \citep{chen2017,malavasi2017,kraljic2018,laigle2018}.
There is also some evidence that low mass late-type galaxies tend to have lower neutral gas content near the filament spine \citep[][]{CroneOdekon2018}, suggesting that galaxies are cut off from their supply of cold gas in this environment.
Massive galaxies on the other hand seem to show evidence of increased cold gas content in the vicinity of filaments \citep[][]{kleiner2017}, providing support for cold mode accretion where galaxies with a large gravitational potential can draw gas from large-scale structure.
The exact mechanism by which galaxies are quenched in the vicinity of the cosmic web filaments is still highly debated. 
It could be driven by the cosmic web detachment \citep[][]{aragon-calvo2016}, where the cut-off of gas supplies near and inside filaments is caused by the turbulent regions inside filaments. 
Another interpretation is based on the cold gas accretion controlled by the filamentary structure \citep[][]{pichon2011,codis2015,laigle2015,welker2017}, where the most efficient helicoidal infall of cold gas is expected in the outskirts of filaments. Our result showing a beyond halo mass effect driving quenching in low-density filaments is first where filaments are split by density. A detailed comparison with a suite of full hydrodynamic simulations should provide insight 
on the physical mechanism of quenching beyond halo mass.

One of the main limitations of this work with regards to constraining
galaxy physics is the lack of small scale information due to fibre
collisions in the observed data. The cross-correlation at small scales is
expected to be especially sensitive to the impact of the cosmic web as well as to the
inter-dependence of different tracers. Measurement at smaller scales
could potentially boost the results in this paper to much higher
significance and we hope to address this regime in future studies.

In summary, the extended halo model is quite powerful in understanding
galaxy properties. It has allowed us to detect a clean signature of
\textsl{1-halo} galactic conformity. It also shows that galaxy
quenching can be explained by a model with halo mass driven quenching in
most parts of the cosmic web (clusters, voids and high density
filaments). But the quenched galaxy fraction in the low-density
filaments is not as predicted by the halo mass quenching model. Our fiducial
\mthod\, model with its variants incorporating galaxy properties beyond 
those due to halo mass 
will thus be fundamental for calibrating models of redshift
space distortions used to obtain cosmological constraints, and to assess the
systematic biases that may arise from the presence of such additional effects.

\section*{Acknowledgements}

SA and JAP are supported by the European Research Council
through the COSFORM Research Grant (\#670193).  We thank Horst Meyerdierks and Eric Tittley for their support with Stacpolly and Cuillin cluster where all of the computing for this project is performed. SA would like to thank Ravi K. Sheth for constructive discussions. We also thank anonymous referee for constructive comments on the initial draft. We thank
the Multi Dark Patchy Team for making their simulations publicly available. This research has made use of NASA's Astrophysics Data System.

Funding for the Sloan Digital Sky Survey IV has been provided by the Alfred P. Sloan Foundation, the U.S. Department of Energy Office of Science, and the Participating Institutions. SDSS-IV acknowledges
support and resources from the Center for High-Performance Computing at
the University of Utah. The SDSS web site is www.sdss.org.

SDSS-IV is managed by the Astrophysical Research Consortium for the 
Participating Institutions of the SDSS Collaboration including the 
Brazilian Participation Group, the Carnegie Institution for Science, 
Carnegie Mellon University, the Chilean Participation Group, the French Participation Group, Harvard-Smithsonian Center for Astrophysics, 
Instituto de Astrof\'isica de Canarias, The Johns Hopkins University, Kavli Institute for the Physics and Mathematics of the Universe (IPMU) / 
University of Tokyo, the Korean Participation Group, Lawrence Berkeley National Laboratory, 
Leibniz Institut f\"ur Astrophysik Potsdam (AIP),  
Max-Planck-Institut f\"ur Astronomie (MPIA Heidelberg), 
Max-Planck-Institut f\"ur Astrophysik (MPA Garching), 
Max-Planck-Institut f\"ur Extraterrestrische Physik (MPE), 
National Astronomical Observatories of China, New Mexico State University, 
New York University, University of Notre Dame, 
Observat\'ario Nacional / MCTI, The Ohio State University, 
Pennsylvania State University, Shanghai Astronomical Observatory, 
United Kingdom Participation Group,
Universidad Nacional Aut\'onoma de M\'exico, University of Arizona, 
University of Colorado Boulder, University of Oxford, University of Portsmouth, 
University of Utah, University of Virginia, University of Washington, University of Wisconsin, 
Vanderbilt University, and Yale University. 

The CosmoSim database used in this paper is a service by the Leibniz-Institute for Astrophysics Potsdam (AIP).
The MultiDark database was developed in cooperation with the Spanish MultiDark Consolider Project CSD2009-00064.

The authors gratefully acknowledge the Gauss Centre for Supercomputing e.V. (www.gauss-centre.eu) and the Partnership for Advanced Supercomputing in Europe (PRACE, www.prace-ri.eu) for funding the MultiDark simulation project by providing computing time on the GCS Supercomputer SuperMUC at Leibniz Supercomputing Centre (LRZ, www.lrz.de).

\section{Data Availability}
All of the observational data used in this paper is available through the SDSS website \url{https://data.sdss.org/sas/dr16/eboss/}. The codes used in this analysis along with instructions are available on \url{https://www.roe.ac.uk/~salam/MTHOD/}.




\bibliography{Master_Shadab}
\bibliographystyle{mnras}


\appendix

\section{Galaxy environment and survey boundaries}
\label{sec:bc-effect}

\begin{figure}
    \centering
    \includegraphics[width=0.48\textwidth]{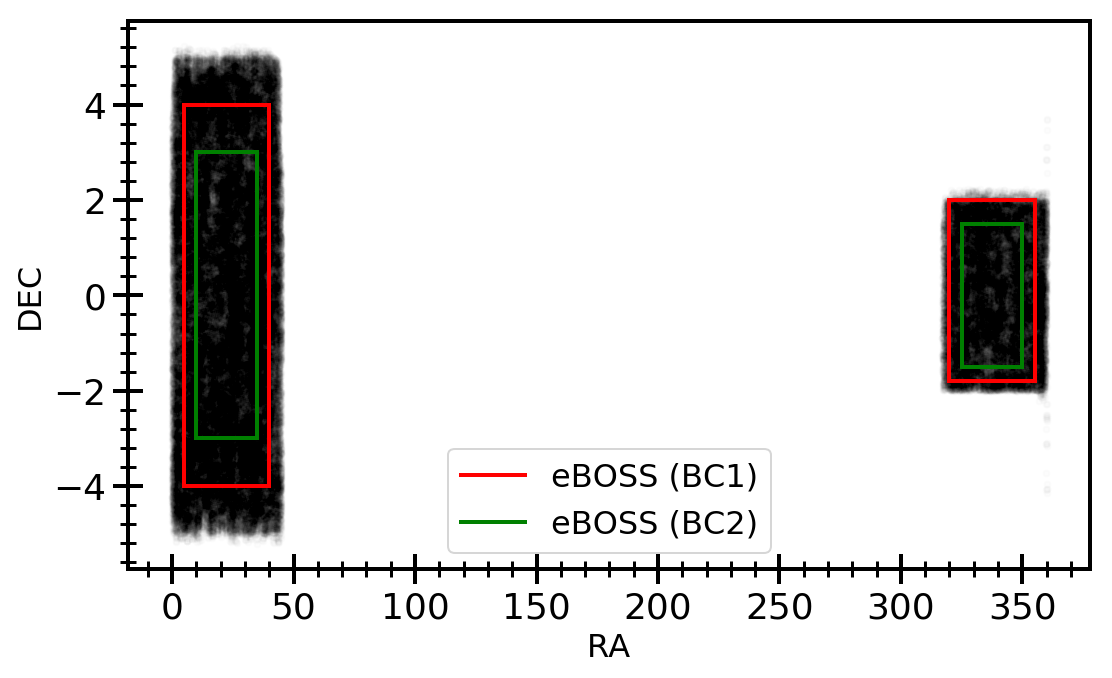}
    \caption{The sky coverage of the full sample is shown in the black and the green and red lines show the boundary cuts used to create the BC1 and BC2 sub-samples respectively.}
    \label{fig:bc}
\end{figure}

\begin{figure}
    \centering
    \includegraphics[width=0.48\textwidth]{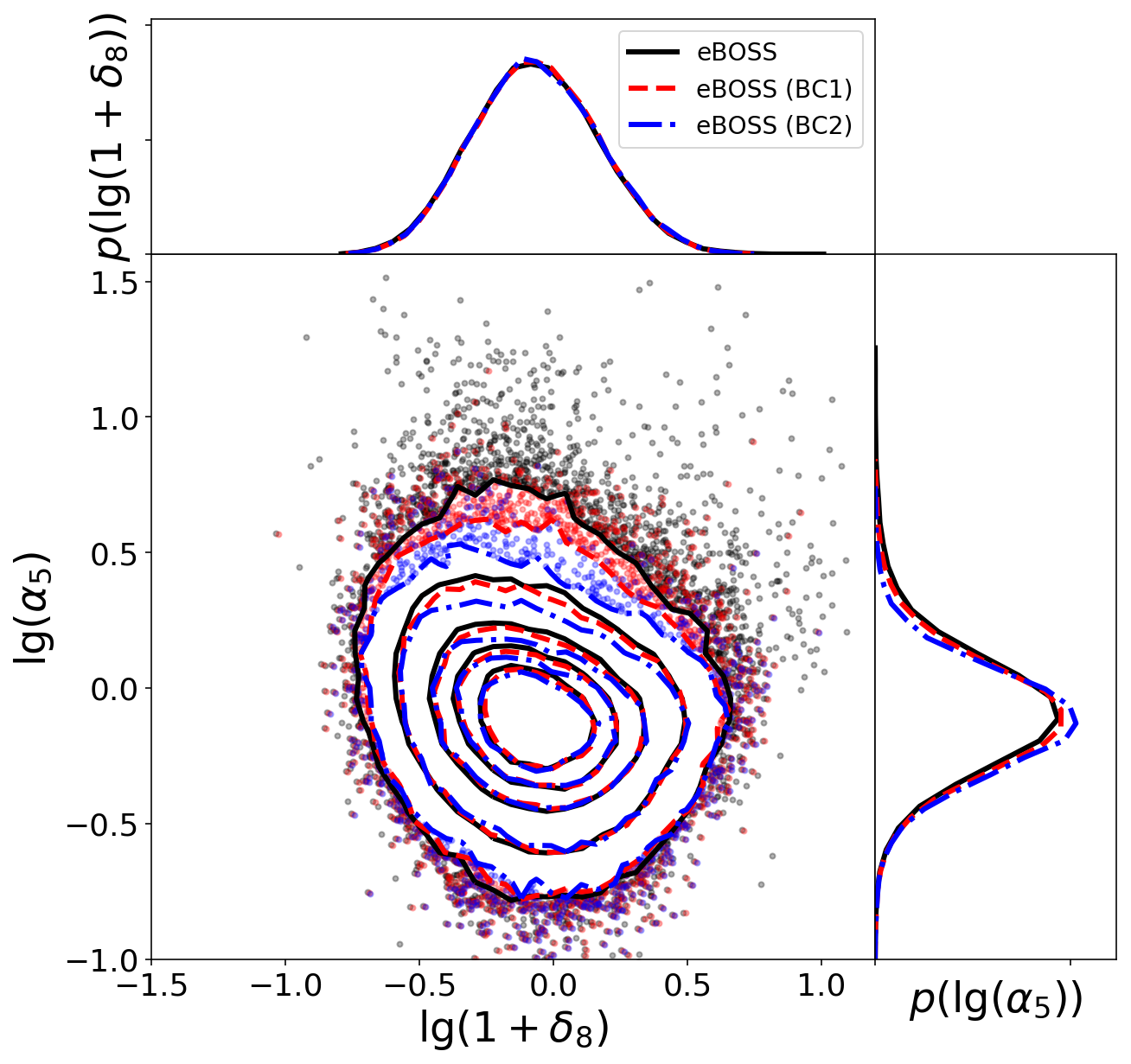}
    \caption{Comparison between the 2D distributions of the full eBOSS galaxy sample and its two sub-samples BC1 and BC2 after applying respective boundary cuts. The two sub-panels on the top and on the right show the marginalised distributions of $\log_{10}(1+\delta_8)$ and $\log_{10}(\alpha_5)$, respectively. The galaxies at the boundary are more populated at high $\alpha_5$ and hence this might reflect boundary effect in our environment calculation.}
    \label{fig:bc-hist2d}
\end{figure}

\begin{figure*}
    \centering
    \includegraphics[width=1.0\textwidth]{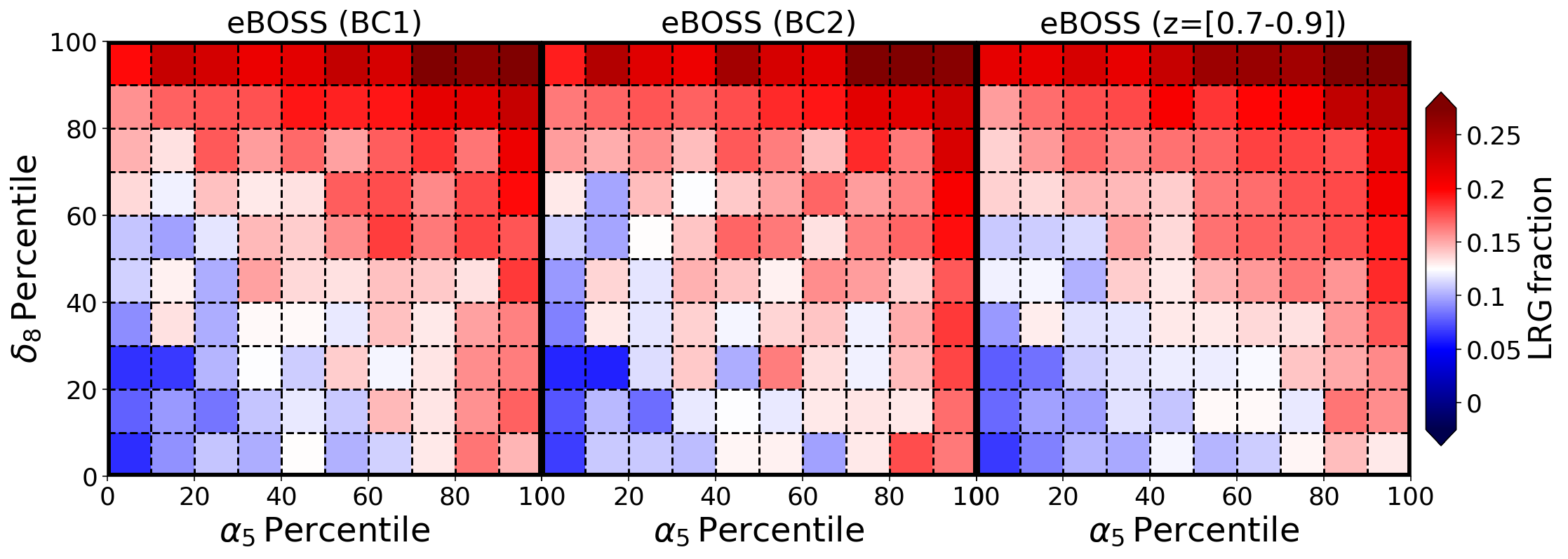}
    \caption{The same as Figure~\ref{fig:delta-alpha-gfrac} but for different boundary cuts. The left, middle and right panels represent the full eBOSS sample, BC1, and BC2 respectively. The overall trend of the LRG fraction in this space is robust against boundary effects.}
    \label{fig:bc-gfrac}
\end{figure*}

For a real survey the estimates of density and tidal field can be
contaminated around the survey boundary. Therefore to investigate such
effects we created various datasets with different degrees of exclusion of
galaxies at the boundary. First we create two sub-samples of our
galaxies by removing galaxies around boundaries called BC1 and
BC2. Figure~\ref{fig:bc} shows the sky distribution of all galaxies
in our sample in black. The two boundary cuts are shown with red and
green lines respectively. For each of these sub-samples we first look
at the distribution of overdensity and tidal anisotropy. Figure~\ref{fig:bc-hist2d} shows the two dimensional and marginalised one
dimensional distribution of galaxies in the $\delta_8-\alpha_5$ space. The
black solid, red dashed and blue dotted-dashed lines show the full
sample, BC1 and BC2 respectively. We find that the distributions of
environmental measures are stable against these boundary effects except regions
with mean $\delta_8$ and high $\alpha_5$. We then look at the LRG fraction
as a function of $\alpha_5$ and $\delta_8$ for each of these
samples. The left and middle panel of Figure~\ref{fig:bc-gfrac} show the LRG fraction for the eBOSS sample with boundary cut of BC1 and BC2 respectively. The overall trend of the LRG fraction seems to be stable against boundary effects. Although the impact of the boundary seems to be small, we use galaxies with BC2 selection for our main results to avoid any possible influence of boundary effects.

We also note that the number density of LRGs drops sharply above redshift 0.9 due to the limits of the telescope, which at high redshift can detect only relative luminous objects. We wanted to assess if our results are robust against such selection in the LRGs. Therefore, we created a sub-sample with boundary cut BC2 and redshift cut between 0.7 and 0.9. The right most panel in Figure~\ref{fig:bc-gfrac} shows the LRG fraction for this sample. We note that the selection effect causing LRG number density to drop-sharply beyond redshift 0.9 has negligible effect in our estimate of the quenching efficiency as a function of tidal environment.

\section{Error in LRG fraction and HOD models}
\label{sec:HOD-env}

\begin{figure*}
    \centering
    \includegraphics[width=1.0\textwidth]{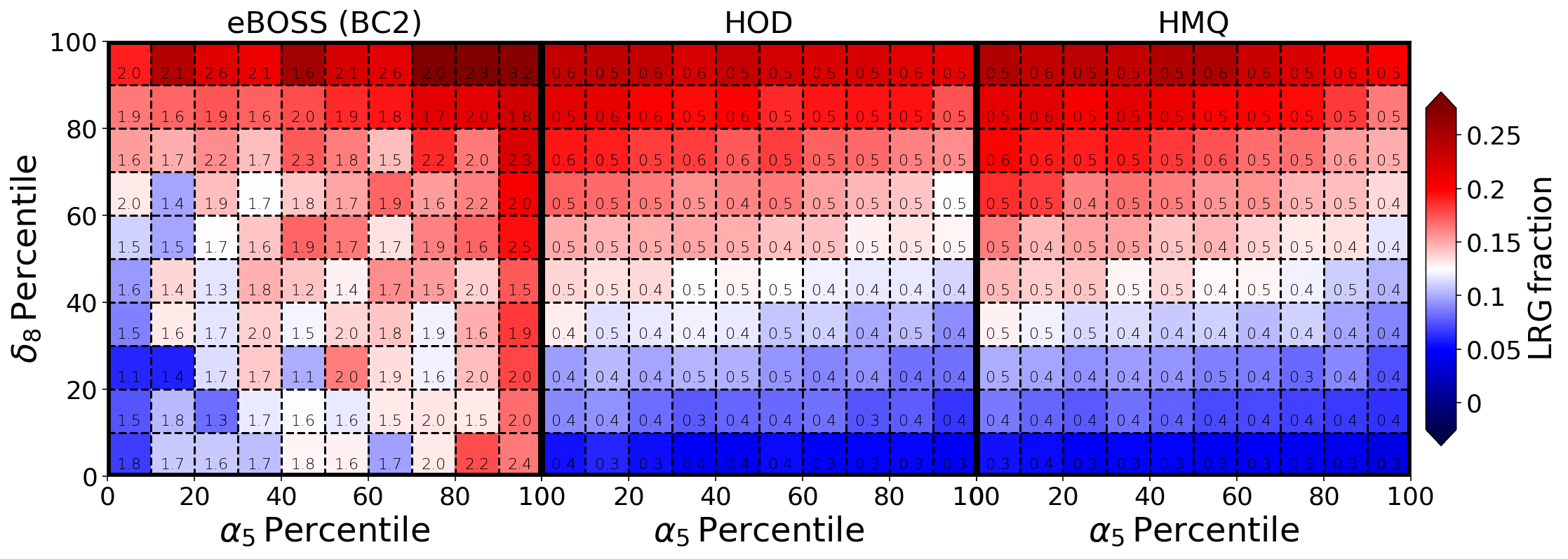}
    \caption{The same as Figure~\ref{fig:delta-alpha-gfrac}, but comparing eBOSS data with the normal and HMQ models including the errors. The left panel shows the eBOSS data with BC2 cuts, the middle panel is for an HOD with no halo mass quenching and the right panel is for the HMQ model. The number displayed in each cell shows 100 times the statistical errors on the LRG fraction.}
    \label{fig:gfrac-HOD-error}
\end{figure*}

Figure ~\ref{fig:gfrac-HOD-error} shows a comparison of the LRG fraction with environment between the standard HOD model and HMQ. We also show 100 times the statistical error on LRG fraction in each cell.


\bsp	
\label{lastpage}
\end{document}